From lab to outbreak: experimental mosquito extrinsic incubation period distributions shape dengue epidemic dynamics

Loisel Léa [1], Sandie Arnoux [1], Gaël Beaunée [1]*, Pauline Ezanno [1]*
[1] Oniris, INRAE, BIOEPAR, Nantes, 44300, France
* These authors contributed equally to this work

## ABSTRACT

Dengue virus transmission models commonly assume an exponential distribution for the mosquito extrinsic incubation period (EIP), potentially oversimplifying biological variability. We developed a stochastic mechanistic dengue transmission model comparing epidemic dynamics under commonly assumed exponential (EXP) versus experimentally derived (ED) EIP distributions. Our results show that using an experimentally derived EIP distribution delays and flattens epidemic peaks, resulting in lower but more prolonged peaks, slightly prolongs crisis durations, and reduces peak intensity compared to the exponential assumption, while outbreak probability remains largely unaffected. These differences are modulated by mosquito mortality and human recovery principally. Incorporating experimentally informed EIP distributions enhances the biological realism of models and may improve predictions of dengue epidemic dynamics, informing more effective vector control strategies and public health responses.

## INTRODUCTION

Over the past two decades, arboviruses have accounted for nearly one-quarter of emerging infections worldwide (1). These viruses are transmitted to vertebrate hosts through the bites of infectious arthropod vectors, such as mosquitoes. Among arboviruses of major public health concern, dengue holds a particularly prominent position due to its broad and expanding global distribution : it is now considered endemic in more than 100 countries across Southeast Asia, Africa, West pacific and America (2). In addition to its worldwide distribution and the predicted increase in the population at risk (3), dengue impose a substantial medical and economic burden, with more than 100 million estimated symptomatic infections per year and over 40 000 deaths (2,4,5).

This ongoing threat posed by dengue, underscores the need for accurate predictive tools to guide timely and effective control strategies. Epidemiological models have been developed for dengue since the 1970s (6,7), and are valuable for understanding and anticipating virus transmission (8), evaluating intervention strategies such as vaccination (9) or vector control (10), and forecasting epidemics across a large range of ecological and climatic contexts (11). However, these models necessarily rely on simplifying assumptions, which can substantially influence their previsions and, consequently, the confidence decision makers place in model outputs (12,13). Therefore, it is crucial to assess the impact of such assumptions on model previsions before using models to support decision-making.

A key assumption in epidemiological models is hidden in the exponential distribution commonly used to represent stage duration during the transmission process (8). This distribution implies a high variance, with most individuals leaving the stage quickly but some remaining in the stage for a long duration. While being parsimonious, requiring only a single parameter, it can be biologically unrealistic. In vector-borne disease models, a critical stage is the vector extrinsic incubation period (EIP), defined as the period between when the vector acquires the virus after biting an infected host and when the virus becomes present in the vector's saliva, making it capable of transmitting the virus to a susceptible host. Vector competence experiments have shown that a log-normal distribution (14) or a beta distribution (15) is more appropriate than an exponential one for modelling the EIP of several arboviruses, including dengue, which questions the use of an exponential distribution for EIP in mechanistic epidemic models.

Indeed, Brand et al. (2016), using a mathematical model of bluetongue virus transmission by *Culicoides*, found that the basic reproduction number ($R_0$) varied according to the theoretical EIP distributions used (16). For dengue, Chowell et al. (2007) first demonstrated that modifying the theoretical distributions of key epidemiological stages, including the EIP in mosquitoes and the intrinsic incubation and infectious periods in humans, influences early epidemic growth (17). Then, Chowell et al. (2013), using a stage-structured vector–host dengue transmission model, showed that the timing and height of epidemic peaks also varied with these theoretical distributions (18). However, these studies relied on theoretical distributions, typically exponential or gamma ones. Therefore, it remains unclear how an experimentally measured EIP distribution would affect epidemic dynamics within a vector–host framework. Comparing an empirical EIP distribution with the commonly used exponential form would clarify under which epidemiological or entomological conditions EIP distribution assumptions meaningfully influence epidemic outcomes.

The present study aims at evaluating and characterizing the impact on dengue epidemics of using an experimentally derived EIP distribution, reflecting the biological variability observed in vector competence experiments, compared to an exponential distribution with the same mean duration. First, we present a stochastic, mechanistic dengue transmission model in which the EIP has either an empirical or an exponential distribution, allowing for a direct comparison of distribution effects. Then, we assess a range of epidemic outcomes, including the probability and type of epidemic, characteristics of the epidemic peak (size, timing, duration, slope), and the onset and duration of crisis phases as defined by symptomatic infection thresholds. Finally, we investigate how the influence of the EIP distribution varies across a range of entomological, demographic, and transmission parameters.

## MATERIALS AND METHODS

**Arbovirus transmission model**

**Model structure and main characteristics**

We developed a mechanistic, stochastic, compartmental model of dengue virus transmission between mosquitoes and humans to assess how the distribution of the within-mosquito state durations affects human epidemic dynamics, including the probability of epidemic emergence that arises from stochastic transmission events (Fig. 1). The model, inspired by Lizarralde-Bejarano and al. (2017) (19), is intentionally parsimonious. However, we modified the formulation of the force of infection to explicitly depend on the relative densities of infectious hosts and vectors. This adjustment ensures that transmission intensity dynamically responds to temporal variations in mosquito population size arising from the introduction of a seasonally varying carrying capacity. It considers a single dengue serotype and two interacting populations. Mosquitoes have an aquatic state ($A_M$), which aggregates eggs, larvae, and pupae, and a female adult state partitioned into **s**usceptible ($S_M$), infected-and-disseminated ($ID_M$), and infectious (transmitter $T_M$), with a total adult mosquito population of size $N_M = S_M + ID_M + T_M$. Humans are distributed into susceptible ($S_H$), exposed ($E_H$)**,** infectious ($I_H$), and recovered ($R_H$) states, with a total human population of size $N_H = S_H + E_H + I_H + R_H$. The time step ($dt$) is of one day, and simulations run over two years (730 days). All transitions between compartments are stochastic: flows between compartments are drawn from binomial distributions, with daily probabilities derived from transition rates as probability = (1-exp$^{-rate.dt}$). Simulations were initialised with a human population of fixed size ($N_H(0)$ = 10 000), with all individuals initially susceptible except for a small number of infectious individuals $I_H(0)$ ranging from 1 to 10 (20)). No mosquito was initially infected, and the initial mosquito population size was scaled proportionally to the human population size according to the mean mosquito-to-human ratio ($\delta$)**.** Parameter definitions and baseline values (drawn from the literature and prior calibration) are summarized in Table 1.

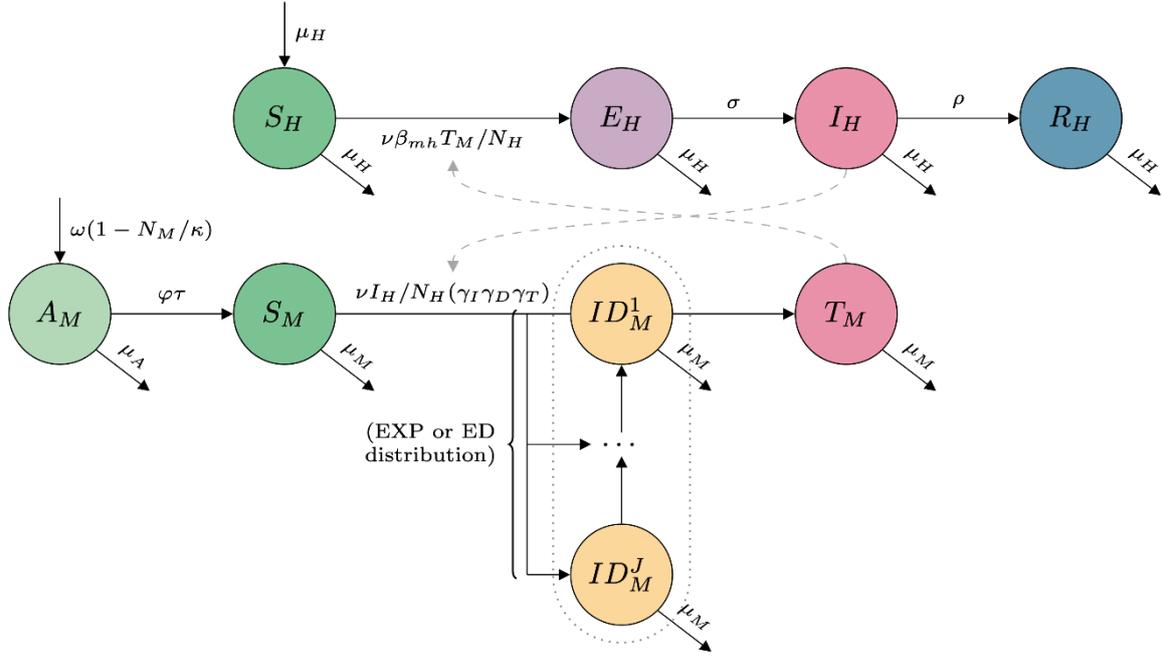

**Figure 1. Model diagram for dengue virus transmission between mosquitoes and humans.**
Mosquitoes distribute among an aquatic state ($A_M$) and states for female adults partitioned into susceptible ($S_M$), infected-and-disseminated ($ID_M$), and infectious (transmitter $T_M$) ones. Total adult mosquito population is of size $N_M = S_M + ID_M + T_M$. Humans distribute among susceptible ($S_H$), exposed ($E_H$), infectious ($I_H$), and recovered ($R_H$) states. Total human population is of size $N_H = S_H + E_H + I_H + R_H$. All transitions between states are drawn from binomial distributions, with daily probability = (1-exp$^{\text{-rate.dt}}$). Refer to Table 1 for parameter definition. Specifically, crossing within-mosquito infection, dissemination and transmission barriers is governed by probabilities ($\gamma_I, \gamma_D, \gamma_T$). Infected mosquitoes progress through successive $ID_M$ sub-compartments according to the assumed extrinsic incubation period (EIP) distribution (experimentally derived (ED) or exponential (EXP)) before becoming infectious ($T_M$), with *J=tmaxID*. Infected mosquitoes whose barriers have not been crossed return to the susceptible state.

### Mosquito epidemiological model (successive transitions to go from $S_M$ to $T_M$)

Instead of using a constant host-to-vector transmission rate followed by an exponential distribution of the EIP, our model explicitly accounts for the variability of the success for the virus of crossing each of the within-mosquito barriers and enables using an experimentally-derived EIP distribution. To do so, it follows 4 steps:
**(1)** it counts the number of $S_M$ mosquitoes biting infectious hosts;
**(2)** it distributes these exposed mosquitoes into subgroups each having a given success probability for the virus to cross the three within-mosquito barriers (infection, dissemination and transmission barriers);
**(3)** it distributes infected mosquitoes (non-infected ones remaining $S_M$) into $ID_M$ sub-compartments ($ID_M^1$ to $ID_M^{tmaxID}$) according to the EIP distribution used (exponential (EXP assumption) or experimentally derived (ED assumption));
**(4)** mosquitoes then deterministically progress across $ID_M^x$ sub-compartments (each of duration of exactly 1 day) until reaching $T_M$.
Therefore, the human to mosquito transmission accounts for the natural variability in within-mosquito viral dynamics across mosquitoes. It remains consistent with experimentally derived within-mosquito transmission dynamics and ensures both the success of infection and the progression speed within the mosquito are jointly consistent. The only difference between the EXP and the ED assumptions thus relies in step 3.

These four steps are detailed hereafter. As a reminder, all model transitions (except the progress among $ID_M$ sub-compartments) are stochastic (proba = (1-exp$^{-rate.dt}$)) and drawn in binomial distribution. For sake of simplicity in model presentation, transitions rates are defined without reminding each time transformation into probability.

**Step (1):** Mosquito exposure

Susceptible mosquitoes ($S_M$) become exposed following bites on infectious humans. The number of newly exposed mosquitoes is determined by a transition rate equal to $\nu \left(\frac{I_H}{N_H}\right)$, where $\nu$ is the biting rate and $N_H$ is the total human population.

**Step (2):** Within-mosquito barriers and inter-mosquito variability

Exposed mosquitoes become infected, disseminated and then transmitter using parameters obtained from the calibration of an intra-vector viral dynamics (IVD) model (15). We previously calibrated this IVD model with experimental data on *Aedes aegypti* infected with dengue virus serotype 3 (DENV-3, 1997/98 strain) (21). We estimated the probabilities for the virus of crossing the three successive within-mosquito barriers, *i.e.*, the probability for an exposed mosquito to be infected ($\gamma_I$), the probability for an infected mosquito to reach disseminated state ($\gamma_D$), and the probability for a disseminated mosquito to become transmitter ($\gamma_T$). Each possible parameter set, made of the triplet ($\gamma_I, \gamma_D, \gamma_T$) together with the two pairs governing the beta distributions of infection ($\alpha_I, \beta_I$) and dissemination ($\alpha_D, \beta_D$) durations (used in the following step), come from a given particle of the posterior inference of the IVD model, which preserves the dependence structure among parameters. To reflect inter-mosquito variability and to scale up from the experimental cohort size ($N_{cohort}$) to the simulated population size, exposed mosquitoes are divided into sub-groups of equal size $N_{cohort}$). For each sub-group, one particle from the IVD model inference is randomly assigned. Within each sub-group, the probability for exposed mosquitoes to stochastically have all their within-mosquito barriers crossed and thus move to the infected-and-disseminated compartment ($ID_M$) is given by the product $\gamma_I \gamma_D \gamma_T$. This approach captures both variability in infection success and progression speed across mosquitoes while remaining consistent with experimental data. For simplicity, the three barriers are treated jointly when transitioning to the *ID* state, since all mosquitoes that reach *ID* will eventually progress to *T*. Allowing mosquitoes to enter *ID* and then revert to *S* if the transmission barrier is not crossed would not add meaningful value to the model.

**Step (3):** EIP distribution (ED vs EXP)

Mosquitoes entering the $ID_M$ compartment are distributed among sub-compartments ($ID_M^1$ to $ID_M^{tmaxID}$), each representing a possible duration in the infected-and-disseminated state. Distribution of duration in $ID_M$ compartment is already defined for each exposed mosquito based on the sub-group it belonged to in step 2. When using an experimentally derived EIP (ED), the sum of the two sampled durations (from the beta distributions of infection ($\alpha_I, \beta_I$) and dissemination ($\alpha_D, \beta_D$)), for respectively the infected and the disseminated states, defines the total time spent by a mosquito in $ID_M$. Under the exponential assumption (EXP), we parameterized an exponential distribution using the sum of the mean durations of the infected and disseminated states derived from the beta distributions, thereby matching the target mean residence time implied by the experimentally inferred parameters. Because the exponential distribution has infinite support, it was truncated at the maximum duration allowed by the experimental observations. We verified that the probability mass beyond this truncation limit was negligible for all parameter sets (Supplementary material, Fig. S1). The reconstructed ED and EXP distributions, and the resulting differences in shape for comparable target mean residence times, are illustrated in Supplementary material, Fig. S2.

**Step (4)**: Progression to $T_M$

Mosquitoes then deterministically progress each day through the $ID_M^x$ sub-compartments until reaching the last one and moving to the $T_M$ compartment. We assume that once this state is reached, mosquitoes remain transmitters for their entire lifespan (22).

### Human epidemiological model

Susceptible humans ($S_H$) become exposed ($E_H$) after being bitten by infectious mosquitoes ($T_M$). The number of newly exposed humans per time step is determined using a rate equal to $\nu \beta_{mh} \left(\frac{T_M}{N_H}\right)$, where $\nu$ is the biting rate, $\beta_{mh}$ is the mosquito-to-human transmission probability, and $N_H$ is the total human population. This formulation reflects the intensity of mosquito–human contacts and the probability of successful infection per bite. Exposed humans ($E_H$) undergo the intrinsic incubation period before becoming infectious ($I_H$) at rate $\sigma$. Infectious humans recover ($R_H$) at rate $\rho$. We assume no return to the susceptible state, as infection with a given dengue serotype is presume to confers lifelong immunity to that serotype (23).

### Population dynamics of mosquitoes and humans

Birth (new individuals entering the aquatic state) and mortality (both in aquatic and adult states) drive mosquito population dynamics. Birth is governed by rate $\omega \left(1 - \frac{N_M}{\kappa(t)}\right)$, where $\omega$ is the oviposition rate, $N_M$ is the total adult female mosquito population and κ(t) is the environmental carrying capacity which limits mosquito population growth. To introduce environmental seasonality over multi-year simulations, κ(t) is modelled as a sinusoidal function, leading to a continuous alternation between favourable and unfavourable periods related to environmental conditions such as food availability and breeding sites: $\kappa(t) = \delta N_H(0)(1 + \eta \sin\left(\frac{2\Pi t}{P}\right))$, where $\delta$ is the mean mosquito-to-human ratio, $N_H(0)$ is the initial human population size, $P$ is the duration of the seasonal cycle expressed in days (365 days for an annual cycle), and $\eta$ is the proportion of variation of the carrying capacity around its mean value, representing the relative amplitude of seasonal variations. $\eta$ ranged from 0 (no seasonality) to 1 (maximum oscillations around the mean), to represent the full range of seasonal dynamics

Adult female mosquitoes emerge from the aquatic state at rate $\varphi \tau$, where $\tau$ is the aquatic-to-adult mosquito transition rate and $\varphi$ is the sexratio. Only females contribute to transmission dynamics and oviposition and thus enter compartment $S_M$. Mosquito mortality is modelled using constant daily rates: $\mu_A$ for the aquatic state and $\mu_M$ for the adult state.

The human population evolves through births and natural deaths, both controlled by the demographic rate $\mu_H$, applied identically to all states irrespective of the infection status (susceptible, exposed, infectious, and recovered).

### Model implementation and sensitivity analysis

The model was implemented in Python. For most parameters, we defined ranges rather than fixed values and generated a large set of parameter combinations to explore model behaviour across a wide spectrum of plausible epidemiological scenarios. Parameter sampling was performed using a Latin Hypercube Sampling (LHS) design, generating a total of 12,000 parameter combinations by varying parameters within their ranges derived from the literature (Table 1). For each parameter combination, we ran 500 stochastic simulations to account for the inherent randomness in the transmission process and to assess the variability and robustness of the resulting model outputs (Supplementary material, Fig S3). The number of replicates was determined from a preliminary analysis of the stability of model outputs percentiles (Supplementary material, Fig. S4).

**Table 1. Model parameters.** Parameters governing within-mosquito infection ($\gamma_I \gamma_D \gamma_T$) are not specified independently but jointly sampled from the IVD model, ensuring consistency with the assumed EIP distribution.

| Parameter | Meaning | Range of values from literature | Ref |
|---|---|---|---|
| **Entomological parameters** | | | |
| $\delta$ | Mean mosquito-to-human ratio | [1-10] | (20,24) |
| $\omega$ | Oviposition rate (day$^{-1}$) | [5-24] | (19,25,26) |
| $\mu_A$ | Aquatic mosquito state mortality rate (day$^{-1}$) | [0.06-0.5] | (19,25,26) |
| $\tau$ | Aquatic to adult mosquito state transition rate (day$^{-1}$) | [0.1-0.2] | (19,25,26) |
| $\mu_M$ | Adult mosquito state mortality rate (day$^{-1}$) | [0.033-0.2] | (6,7,19,24–27) |
| $\nu$ | Biting rate (day$^{-1}$) | [0.2-0.7] | (6,7,19,26,28) |
| $\varphi$ | Mosquito sex ratio (fraction of female mosquitoes produced during aquatic to adult transition) | 0.5 | (19) |
| $\eta$ | Relative amplitude (proportion) of seasonal variations in mosquito carrying capacity around its mean | [0-1] | / |
| **Human parameters** | | | |
| $\mu_H$ | Human demography rate (day$^{-1}$) | 0.000046 | (6) |
| **Arbovirus transmission parameters** | | | |
| $\gamma_I \gamma_D \gamma_T$ | Probability for exposed mosquitoes to have all their within-mosquito barriers crossed | [0.0639574-0.8028093] | (15) |
| $\beta_{mh}$ | Mosquito to human transmission probability | [0.1-0.8] | (6,26) |
| $\rho$ | Human recovery rate (day$^{-1}$) | [0.07-0.33] | (6) |
| $\sigma$ | Exposed to infectious human transition rate (day$^{-1}$) | [0.1-0.25] | (29) |

**Human epidemic indicators**

We analysed the characteristics of human epidemics using a set of model outputs (thereafter named epidemic indicators, Table 2 and Supplementary material, Fig. S5). We designed these indicators to capture three key dimensions of epidemic dynamics: (i) magnitude (total number of human infections, epidemic peak size, epidemic type and crisis occurrence), (ii) timing (peak and crisis dates, peak slope), and (iii) duration (peak and crisis durations). Together, these indicators allow for a comprehensive assessment of the amplitude, speed, and temporal profile of human outbreaks as generated by the model. Epidemics were classified according to their magnitude and severity based on the proportion of the human population infected during the simulation. We considered an epidemic has occurred when the final cumulative number of human infections ($I_{Hcum}$) exceeded twice the initial number of infectious individuals ($I_{Hcum} > 2 \times I_H(0)$). Among epidemics, we defined as large epidemics those with $I_{Hcum} > 1\%$ of the initial human population ($I_{Hcum} > 100$ for a population of 10 000 (20)). Among large epidemics, we defined as crises, situations in which at least 500 individuals were simultaneously infectious ($I_H \geq 500$). This threshold corresponds to 100 simultaneous symptomatic cases in a population of 10,000 people (1% of the total population), assuming that 80% of dengue infections are asymptomatic (30,31). Such a symptomatic burden is here representative of a severe epidemic situation.

**Table 2. Summary of indicators used to characterize human epidemic dynamics.**

| Output name | Description | Purpose |
|---|---|---|
| **Number of epidemics** | Number of simulations in which the final cumulative number of human infections exceeds twice the number of initially infectious individuals ($I_{Hcum} > 2 \times I_H(0)$). | Definition based on $R_0$: an epidemic is considered to occur if each individual initially infected infects more than one other individual. |
| **Total epidemics size** | Final cumulative number of human infections over the two-year simulation period ($I_{Hcum}$). | Quantifies the overall scale of the epidemic and its potential severity. |
| **Number of small / large epidemics** | Among epidemics, distinction between those where the total epidemic size is below or above 1% of the total human population size (i.e. $I_{Hcum} < 100$ or $I_{Hcum} \geq 100$ for a population of 10,000). | Differentiates small outbreaks from large epidemics based on a scalable threshold relative to population size and response capacity. |
| **Date of epidemic peak (small / large epidemics)** | Day at which the number of infectious humans at each time step ($I_H$) reaches its maximum value. | Assesses the date when the epidemic peaks, useful to adapt the available management resources and measures over time. |
| **Size of epidemic peak (small / large epidemics)** | Maximum value for $I_H = (I_{H\,max})$. | Assesses the magnitude of the epidemic peak to quantify the maximal pressure exerted on the population and assess potential needs for intervention and resource allocation. |
| **Epidemic peak start, duration, and end (small / large epidemics)** | Time interval between the first and last day during which the number of infectious humans ($I_H$) exceeds 10% of its maximum value. | Assesses the duration of the epidemic peak to adapt the available management resources and measures over time. |
| **Epidemic peak growth rate (peak slope)** | $$\frac{I_{H\,max} - I_{H\,10\%}}{t(I_{H\,max}) - t(I_{H\,10\%})}$$ Average daily increase in the number of infectious individuals between 10% of the epidemic peak size and the peak. | Measures the speed of epidemic growth. |
| **Number of crises** | Number of simulations in which the number of infectious humans reaches at least 500 at any time ($I_H \geq 500$; threshold equivalent to 100 symptomatic cases assuming 80% of infections are asymptomatic (30)). | Assesses how many simulations lead to a crisis, the definition of which can be adapted according to the human and material resources available to manage the epidemic. |
| **Crisis start, end, and duration (large epidemics)** | Time interval between the first and last day during which the number of infectious humans is at least 500 ($I_H \geq 500$). | Quantifies the timing and duration of healthcare system stress. |

**Comparison between exponential (EXP) and experimentally derived (ED) distribution for extrinsic incubation period (EIP)**

To assess how the assumed distribution of the mosquito extrinsic incubation period (EIP) influences epidemic dynamics in humans, we compared the epidemic indicators obtained when assuming an exponential (EXP) vs. an experimentally derived (ED) EIP distribution, across all parameter combinations. Because of model stochasticity, each parameter combination is associated to two distributions (one for EXP, one for ED) for each epidemic indicator. We used complementary statistics to compare these distributions between EIP distribution assumptions and to evaluate how the magnitude of their differences varied among parameter combinations. First, we applied a Wilcoxon rank-sum test (32) to compare the two distributions for each of the epidemic indicators in each of the parameter combination. Second, we calculated the proportion of tests yielding significant differences (*p-value* < 0.05) to quantify how often the two EIP distribution assumptions led to statistically distinct epidemic indicators. In addition, we computed the Hodges–Lehmann (HL) estimator (33). HL is the median of all pairwise differences between values from the two distributions assumptions and is expressed in the same unit as the corresponding epidemic indicator (e.g., number of infected individuals for peak size, days for peak timing). The sign of the HL value indicates the direction of the difference (positive or negative), allowing direct interpretation of which EIP distribution assumption leads to the highest epidemic indicator. We also quantified the differences in distribution assumption shape using the one-dimensional Wasserstein distance (34). This metric captures the overall difference between two distributions by measuring how much and how far values from one distribution would need to be shifted to match the other.

To identify which parameters most influenced the variation of the HL estimator and the Wasserstein distance, we performed an ANOVA relying on linear models fitted to the LHS design, including all main effects and pairwise interactions. All input parameters were centred and scaled prior to model fitting to ensure numerical stability and harmonise their ranges obtained from literature (Table 1). The relative importance of each parameter was assessed using a Type-II ANOVA, which quantifies their conditional contribution to the variability of the HL estimator and the Wasserstein distance accounting for the others parameters in the model (35). Contributions of each parameter were summarized at the parameter level to account for both main effects and interactions. Doing so, we identified parameters influencing differences between the EXP and ED assumptions, after which we examined how the HL estimator and the Wasserstein distance varied with these key parameters.

**Software and computational resources details**

All computations were run on the MIGALE computing cluster. Model was run on python. All epidemics indicator analyses were conducted in R. Wilcoxon rank-sum test was applied with the wilcox.test() function in R (*stats* package). The one-dimensional Wasserstein distance was computed with the wasserstein1d() function from the *transport* package. Type-II ANOVA was performed on linear models including all main effects and pairwise interactions, using the Anova () function from the car package in R.

**Code and data availability**

The code used to perform the analyses and generate the results presented in this study is available from the project GitHub repository: human_mosquito_arbov_pop_and_within_mosquito available at: https://forge.inrae.fr/dynamo/vbd/human_mosquito_arbov_pop_and_within_mosquito . Input data required to reproduce the main analyses are provided in the repository.

## RESULTS

**EIP distribution shapes intensity, timing, and duration of epidemic peaks**

Using an exponential (EXP) or an experimentally derived (ED) distribution for modelling the mosquito extrinsic incubation period (EIP) duration (i.e. the infected-and-disseminated state duration) influenced the intensity (size), timing (date), velocity (slope), and duration of the epidemic peak in humans (Fig. 2; see Table 2 for indicators' definitions). The EXP assumption usually produced higher, earlier, and sharper peaks, albeit of shorter duration, than the ED assumption. This pattern was particularly consistent for large epidemics with crisis, for which peak-related indicators (size, timing, duration and slope) all followed the same trend (Fig. 2A-D). The Hodges–Lehmann (HL) estimators across parameter combinations ranged from -1188.45 to -242 infectious individuals for peak size (5th–95th percentile range), with a median of -478 infectious individuals (Fig. 2Ai). For peak slope, the HL values ranged from -62.46 to -8.25 (5th–95th percentile range) with a median of -25.36 (Fig. 2Ci). For peak date, HL values ranged from 23 to 89 days (5th–95th percentile range), with a median of 41 days (Fig. 2Bi). For all the three indicators, 100% of parameter combinations showed significant differences between EXP and ED assumptions (Wilcoxon rank-sum test, p-value < 0.05; Figs. 2Ai-Ci). The corresponding violin plots illustrate these differences, with lower medians under ED for peak size and slope, and higher medians for peak date (Figs. 2Aii-Cii). For peak duration (Fig. 2D), all HL estimators were positive, ranging from 9 to 93 days (5th–95th percentile range) with a median of 24 days (Fig. 2Di), showing that peaks tended to last slightly longer under ED assumption. We obtained similar conclusions using the Wasserstein distance for large epidemics with crisis (Supplementary Fig. S6) and observed consistent patterns for both HL estimators and Wasserstein distances in large epidemics without crisis ($I_H$ < 500), although with greater dispersion (Supplementary Figs. S7–S8).

**EIP distribution affects timing and duration of epidemic crises**

Beyond peak characteristics, the EIP distribution also affected the timing and persistence of epidemic crises (Fig. 3). Assuming an EXP EIP distribution almost always produced earlier and slightly shorter crises than using an ED EIP distribution. For the crisis onset (Fig. 3A), 100 % of the parameter combinations displayed a significant difference between EXP and ED (Wilcoxon rank-sum test, p < 0.05). The distribution of Hodges–Lehmann (HL) estimators across parameter combinations ranged from 19 to 91 days (5th–95th percentile range), with a median of 38 days (Fig. 3Ai), indicating that crises consistently began earlier under the EXP assumption. The corresponding violin plots (Fig. 3Aii) illustrate this trend, showing a temporal shift toward earlier crisis onset under the EXP assumption.
For the crisis duration (Fig. 3B), differences between EXP and ED were smaller but still significant, with 99% of the parameter combinations showing a significant difference (p-value < 0.05). HL estimators ranged from -21 to 16 days (5th–95th percentile range), with a median of 5 day (Fig. 3Bi), indicating that crises tended to last slightly longer under the ED distribution. The violin plots (Fig. 3Bii) illustrate this subtle extension in crisis duration. Similar conclusions were obtained when using the Wasserstein distance for large epidemics with crisis (Supplementary Material, Fig. S6).

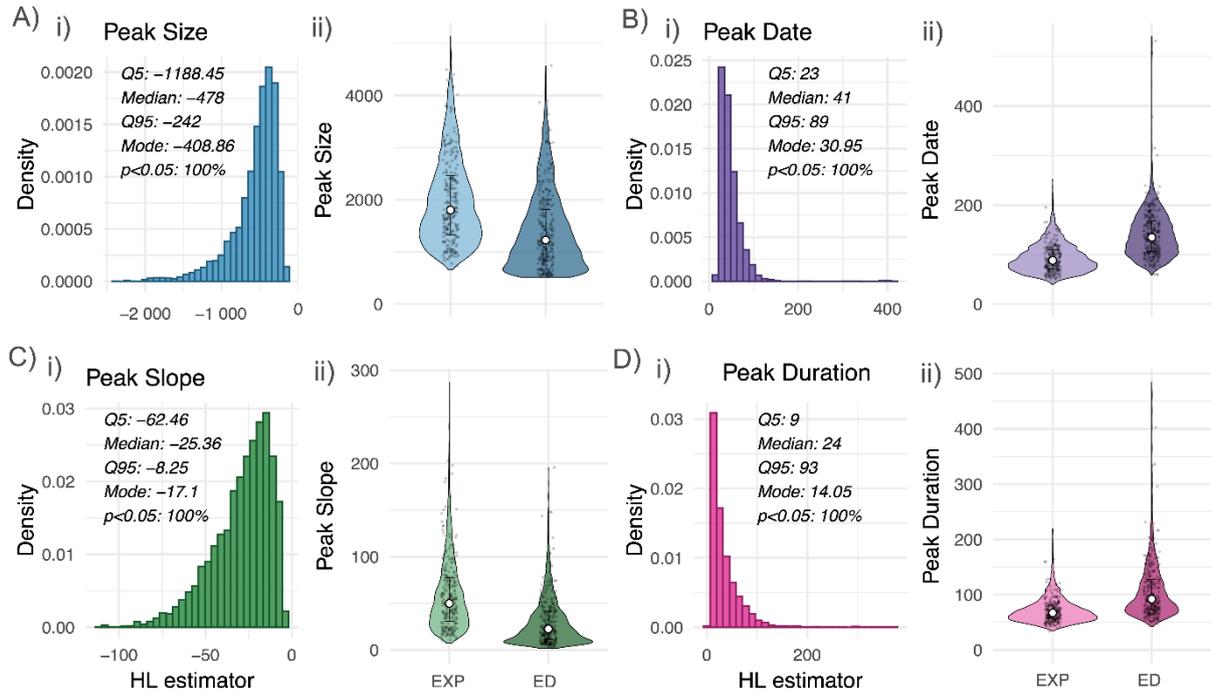

**Figure 2. Differences in peak-related epidemic indicators (large epidemics with crisis) when using an exponential (EXP) *vs.* an experimentally derived (ED) extrinsic incubation period (EIP) distribution.** (A) Peak size ($I_{H_{max}}$, maximum number of concurrently infectious humans), (B) peak date (time of $I_{H_{max}}$), (C) peak slope (average daily increase in $I_H$ between 10 % of the peak size and the peak size), and (D) peak duration (time interval between the first and last day when $I_H$ exceeded 10 % of peak size). Each epidemic indicator is summarized by: (i) a histogram of the Hodges-Lehmann (HL) estimator of the difference between epidemic indicator for ED and EXP EIP computed per parameter combination* (negative value: larger indicator with EXP EIP); (ii) violin plots showing the distribution of the medians of epidemic indicators for each parameter combination for each assumption (small black dots: sample of medians (500 max); horizontal bar: Q25-Q75 interquartile range; white–black circle: median). Insets in (i) report quantiles, mode, and the proportion of parameter combinations with statistically significant differences between EXP and ED assumptions (Wilcoxon rank-sum test, p-value < 0.05). *Parameter combinations were generated using a Latin Hypercube Sampling (LHS) design, in which parameter values were simultaneously varied across their literature-based ranges to produce 12,000 combinations. Analysis is restricted to parameter combinations that produced at least 100 large epidemics with crisis under each assumption (i.e. total number of human infections ≥ 100 and at least 500 concurrently infectious individuals), resulting in 3952 parameter combinations used.*

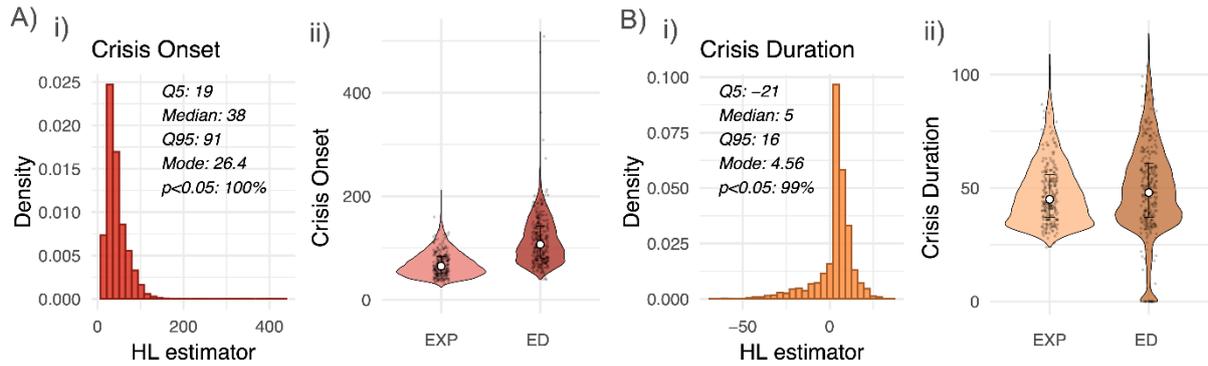

**Figure 3. Differences in crisis-related epidemic indicators (large epidemics with crisis) when assuming an exponential (EXP) *vs.* an experimentally derived (ED) extrinsic incubation period (EIP) distribution.** Panels: (A) Crisis onset (first day when the number of concurrently infectious humans reached the "crisis" threshold, $I_H \geq 500$), (B) crisis duration (time interval between the first and last day when $I_H \geq 500$). Each outcome is summarized by: (i) a histogram of the Hodges–Lehmann (HL) estimator of the difference between epidemic indicator for ED and EXP computed per parameter combination* (negative value: larger indicator under EXP assumption); (ii) violin plots showing the distribution of the medians of epidemic indicators for each parameter combination under each assumption (small black dots: sample of medians (500 max); horizontal bar: Q25-Q75 interquartile range; white–black circle: median). Insets in (i) report quantiles, mode, and the proportion of parameter combinations with statistically significant differences between assumptions (Wilcoxon rank-sum test, p-value < 0.05). *Parameter combinations were generated using a Latin Hypercube Sampling (LHS) design, in which parameter values were simultaneously varied across their literature-based ranges to produce 12,000 combinations. Analysis was restricted to parameter combinations that produced at least 100 large epidemics with crisis under each assumption (i.e. total number of human infections ≥ 100 and at least 500 concurrently infectious individuals at any time), resulting in 3952 parameter combinations used.*

**EIP distribution has small and context-dependent effects on total epidemic size**

The total epidemic size (final cumulative number of human infections over the two-year simulation period) for large epidemics was affected differently by the EIP distribution depending on whether a crisis occurred or not (Fig. 4). Indeed, for large epidemics with crisis, the distribution of Hodges-Lehmann (HL) estimators across parameter combinations were strongly concentrated around 0, with a mode of 11.31 human infections (Fig. 4Ai), indicating that differences between the EXP and ED assumptions were generally small. When differences occurred, they usually reflected larger epidemic sizes under the EXP assumption, with HL values ranging from -1412 to 18 human infections (5th–95th percentile range). The violin plots (Fig. 4Aii) illustrate this tendency. In contrast, differences were more pronounced for large epidemics without crisis (Fig. 4B), with a broader HL distribution (-6863.2 to -308.5 human infections (5th–95th percentile range); Fig. 4Bi) and violin plots (Fig. 4Bii) showing higher median epidemic sizes under the EXP assumption. This stronger dispersion likely arises from the wider heterogeneity of epidemic profiles included in the "large without crisis" category, which encompasses a broad spectrum of epidemic magnitudes.

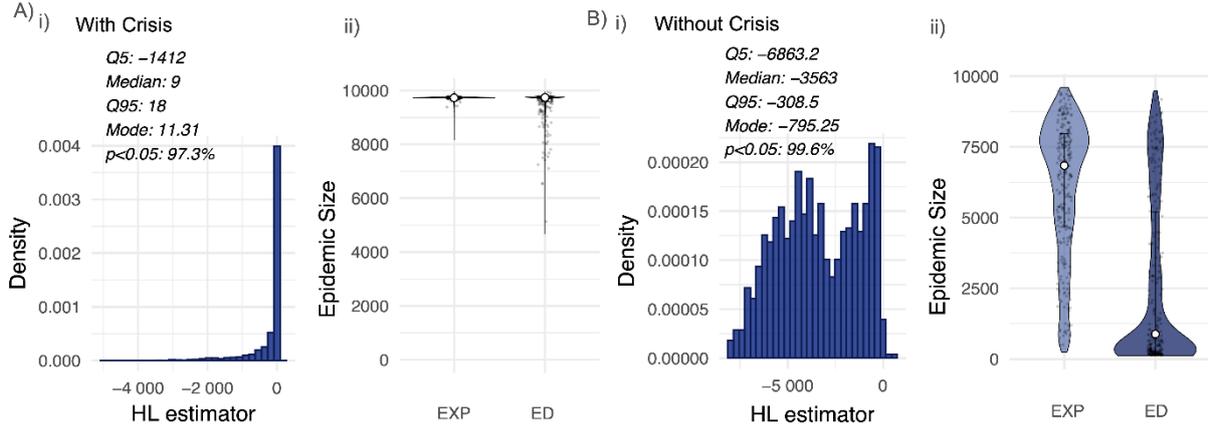

**Figure 4. Differences in total epidemic size (large epidemics with and without crisis) when assuming an exponential (EXP) *vs.* an experimentally derived (ED) extrinsic incubation period (EIP) distribution**. Panel A) shows figures for large epidemics with crisis and B) for large epidemics without crisis. (i) histogram of the Hodges–Lehmann (HL) estimator of the difference between epidemic size under ED and EXP assumptions computed per parameter combinations* (negative value: larger indicator under EXP assumption); (ii) violin plots showing the distribution of the medians of the epidemic size for each parameter combination under each assumption (small black dots: sample of medians (500 max); horizontal bar: Q25-Q75 interquartile range; white–black circle: median). Insets in (i) report quantiles, mode, and the proportion of parameter combinations with statistically significant differences between assumptions (Wilcoxon rank-sum test, p-value < 0.05).
*Parameter combinations were generated using a Latin Hypercube Sampling (LHS) design, in which parameter values were simultaneously varied across their literature-based ranges to produce 12,000 combinations. Analysis was restricted: for (A) to parameter combinations that produced at least 100 large epidemics with crisis under each assumption (i.e. total number of human infections ≥ 100 and at least 500 concurrently infectious individuals at any time), resulting in 3952 parameter combinations used; for (B) to parameter combinations that produced large epidemics without crisis under each assumption (i.e. total number of human infections ≥ 100 and fewer than 500 concurrently infectious individuals at any time), resulting in 942 parameter combinations used.*

**EIP distribution has limited influence on epidemic occurrence**

In contrast to the strong effects on epidemic dynamics, the EIP distribution had only a limited influence on epidemic occurrence (Fig. 5). For each parameter combination, the probability of occurrence was defined as the fraction of stochastic simulations resulting in a given epidemic type: *epidemics* ($I_{Hcum} > 2 \times I_H(0)$), *small epidemics* ($I_{Hcum} < 100$), *large epidemics* ($I_{Hcum} \geq 100$), and among the latter, *large epidemics with crisis* ($I_H \geq 500$ at any time) or *without crisis* ($I_H < 500$ at all times). Across all categories, the distributions of differences in epidemic occurrence probabilities (ED − EXP) were sharply peaked at zero (Fig. 5i–iii and Fig. S9), indicating that the assumed EIP distribution had barely any effect on the probability of epidemic occurrence. When differences occurred, they consistently reflected a higher probability of epidemic occurrence under the EXP assumption.

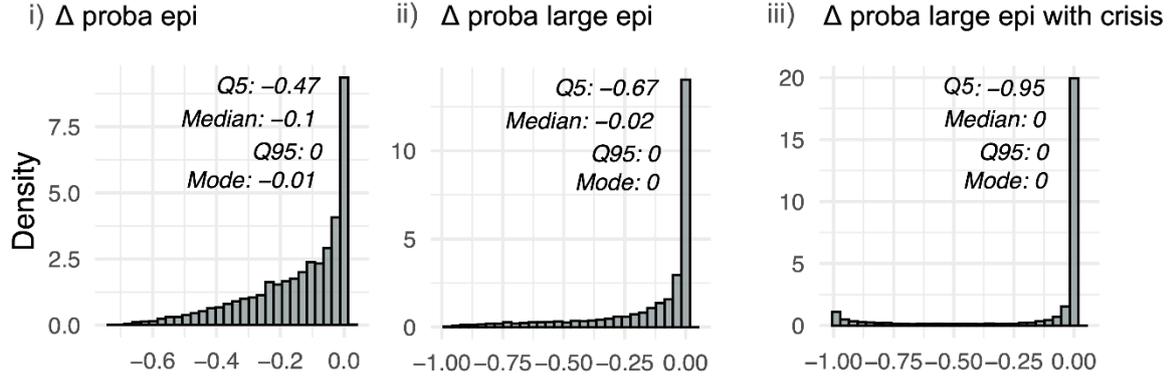

**Figure 5. Differences in probability of epidemic type when assuming an exponential (EXP) *vs.* an experimentally derived (ED) extrinsic incubation period (EIP) distribution**. Histogram of the differences (ED – EXP) between probability that a parameter combination* leads to i) an epidemic (final cumulative number of human infections ($I_{Hcum}$) exceeded twice the initial number of infectious individuals ($I_{Hcum} > 2 \times I_H(0)$)), ii) a large epidemic ($I_{Hcum} \geq 100$), iii) a large epidemic with crisis ($I_H \geq 500$). Insets report quantiles and mode. *Parameter combinations were generated using a Latin Hypercube Sampling (LHS) design, in which model parameters were simultaneously varied across their literature-based ranges to produce a total of 12,000 parameter combinations.*

**Differences between EXP and ED assumptions vary across parameter values for large epidemic with crisis**

To assess how variations in model parameters modulate the differences between the EXP and ED assumptions, we analysed the variations of the Hodges–Lehmann (HL) estimators across the Latin Hypercube Sampling (LHS) design. For each epidemic indicator, parameter contributions were summarized using ANOVA-based relative contributions and displayed as a heatmap (Fig. 6). Because colour intensities were normalized within each epidemic indicator, the heatmap should be interpreted column by column to the parameters most strongly associated with HL variability for a given indicator. Several consistent patterns emerged for large epidemics with crisis. Across most epidemic indicators, the adult mosquito mortality rate ($\mu_M$) showed the largest relative contribution to HL variability, highlighting the prominent role of vector survival in shaping differences between the EXP and ED assumptions within the explored parameter space. The human recovery rate ($\rho$) also contributed substantially, particularly for crisis duration and peak slope, indicating that host infectious period modulates differences in epidemic growth and persistence. Parameters governing transmission intensity, biting rate ($\nu$), mosquito-to-human transmission probability ($\beta_{mh}$) and the mean mosquito-to-human ratio ($\delta$), were especially influential for crisis duration, peak duration and peak slope. In contrast, infection initial conditions ($I_H(0)$), exposed to infectious human rate ($\sigma$) and other entomological rates (oviposition rate ($\omega$), aquatic mortality rate ($\mu_A$), aquatic-to-adult transition rate ($\tau$), and amplitude of varying carrying capacity ($\eta$)) showed low contributions across all epidemic indicators. Similar qualitative patterns were observed for Wasserstein-based divergence measures and for large epidemics without crisis (Supplementary Material., Figs S10-S12). However, for large epidemics without crisis, parameter effects were less clearly structured, with stronger interaction effects, limiting the dominance of individual main effects.

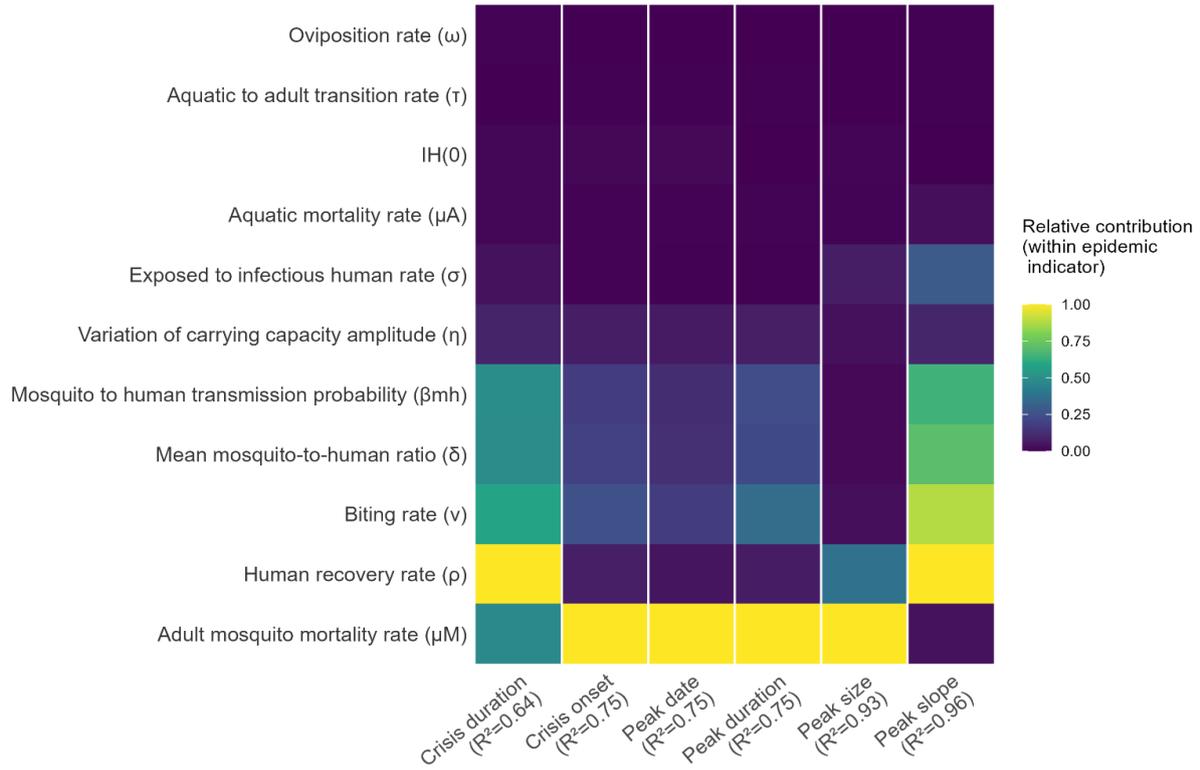

**Figure 6. Relative contribution of model parameters to differences between EXP and ED assumptions for large epidemics with crisis estimated with the Hodges-Lehmann estimator.** Heatmap of ANOVA-based relative contributions of model parameters to the variability of the Hodges–Lehmann (HL) estimators, computed across the Latin Hypercube Sampling (LHS) design restricted to parameter combinations that generated at least 100 large epidemics with crisis. For each epidemic indicator, parameter contributions were derived from second-order linear models (including main effects and pairwise interactions) and aggregated per parameter. Color intensities represent the relative contribution of each parameter to the variability of HL for a given epidemic indicator, rescaled to the [0,1] range within each indicator.

To explore how key parameters modulate the differences between EXP and ED assumptions, we examined how the HL estimators varied with each of them (Fig. 7). Across parameters, adult mosquito mortality rate ($\mu_M$) showed the clearest effect: increasing $\mu_M$ was associated with larger differences between EXP and ED, particularly for peak timing and peak size. In contrast, higher human recovery rates ($\rho$) were associated with reduced differences between EIP distribution especially for peak size and peak slope. The biting rate ($\nu$), mean mosquito-to-human ratio ($\delta$) and the mosquito-to-human transmission probability ($\beta_{mh}$) showed weaker and more variable relationships with HL estimators. Although slight trends suggested that higher transmission intensity may marginally enhance differences between EIP distributions for peak slope, their overall influence appeared limited compared to $\mu_M$ and $\rho$. Similar patterns were observed for crisis onset and crisis duration (Fig. S13). Trends for large epidemics without crisis were more difficult to identify due to greater dispersion of HL estimators, likely reflecting a stronger role of parameter interactions and structural heterogeneity in this epidemic category (Figs. S14). Consequently, clear relationships between individual parameters and HL estimators could only be identified for large epidemics with crisis.

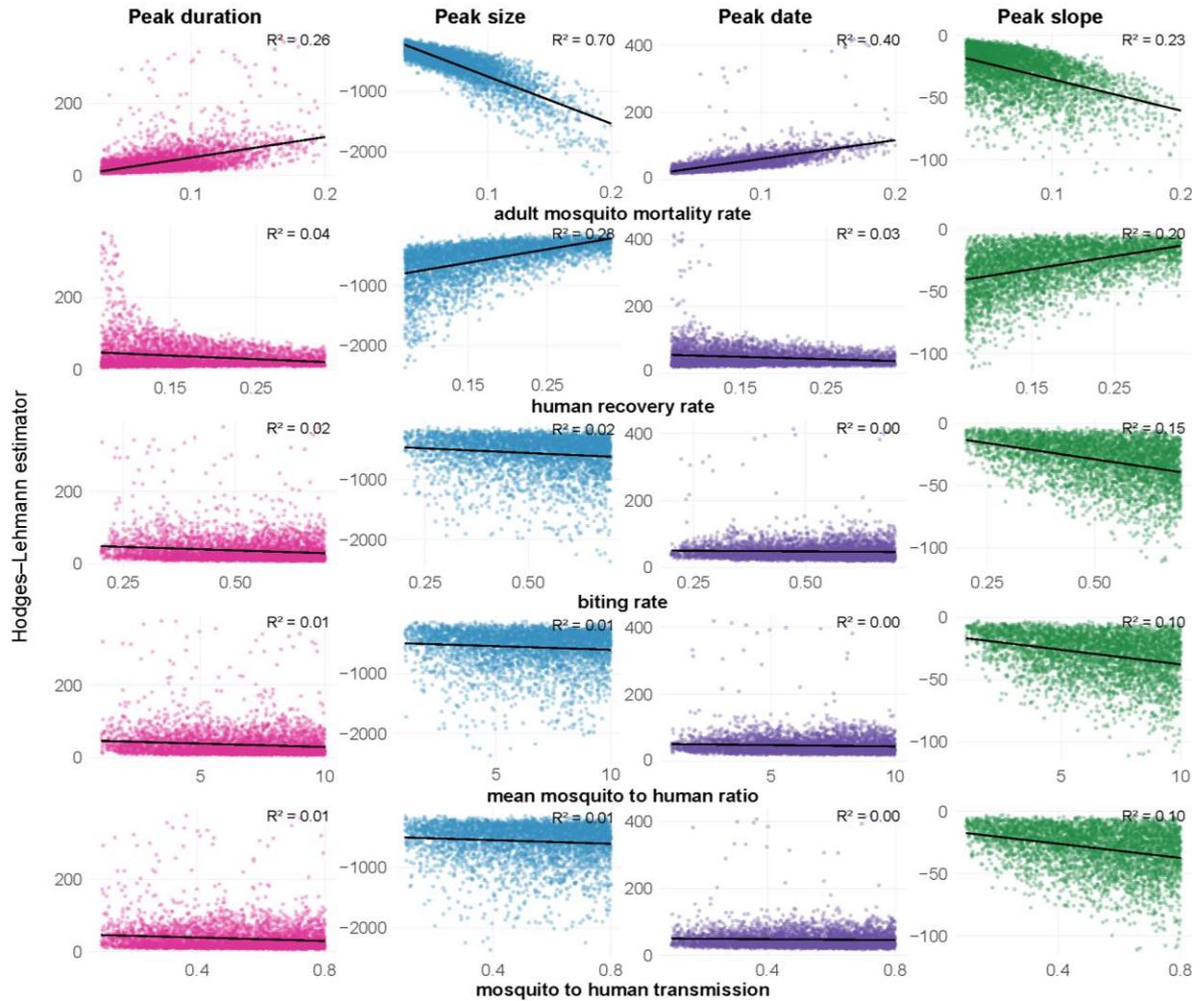

**Figure 7. Impact of key parameters on differences between the EXP and ED assumptions for large epidemics with crisis.** Each panel shows the marginal variation of the Hodges–Lehmann (HL) estimator with respect to one key parameter, while all other parameters vary across the Latin Hypercube Sampling design restricted to parameter combinations that generated at least 100 large epidemics with crisis. Columns correspond to epidemic indicators: (1) peak duration (time interval between the first and last day when $I_H$ exceeded 10% of the peak value, (2) peak size ($I_{H_{max}}$, maximum number of concurrently infectious humans), (3) peak date (time of $I_{H_{max}}$), and (4) peak slope (average daily increase in $I_H$ between 10 % of the peak size and the peak size. Rows correspond to parameters: (1) adult mosquito mortality rate ($\mu_M$), (2) human recovery rate ($\rho$), (3) biting rate ($\nu$), (4) mean mosquito-to-human ratio ($\delta$) and (5) mosquito-to-human transmission probability ($\beta_{mh}$). Points represent individual HL estimators, and black lines indicate average linear trends shown for visual guidance only and not intended for quantitative inference.

**DISCUSSION**

Using a stochastic vector-host dengue transmission modelling framework, we evaluated the impact of alternative assumptions for the mosquito extrinsic incubation period (EIP), using either a theoretical exponential distribution (EXP) or an experimentally derived distribution (ED). We showed that the choice of the mosquito EIP distribution has a measurable impact on epidemic dynamics in humans. Using an experimentally derived EIP distribution affected the timing, magnitude and duration of the epidemic peak, as well as the timing of epidemic crises, while having more limited effects on crisis duration. It slightly influenced the total epidemic size and had no detectable effect on the probability of outbreak occurrence. These effects varied with parameter values for the mosquito death rate, the human recovery rate, the biting rate, the mosquito-to-human transmission probability and the mean mosquito-to-human ratio. This highlights that the choice of the EIP distribution should be made carefully, depending on the epidemic's indicators used and the epidemiological and entomological contexts.

Beyond quantifying the effect of the EIP distribution on epidemic dynamics, our work also provides a framework to incorporate intra-vector experimental data into vector–host transmission models. This approach allows for the joint calibration and inclusion of both the EIP distribution and the human-to-mosquito transmission parameter in the vector-host model, two parameters that are often difficult to estimate due to the wide range of values reported in the literature (6,7).

Overall, assuming an exponential EIP distribution resulted in earlier and higher epidemic peaks and earlier crisis onset, whereas using an experimentally derived distribution produced flatter, later, and longer epidemic peaks, as well as slightly longer crisis durations. These findings are consistent with previous modelling works comparing exponential and other theoretical distributions (17,18). Our study extends these findings by directly comparing the exponential distribution to an experimentally derived distribution, which more closely reflects the biological variability of mosquito infection dynamics. Our results also indicate that using an EXP or ED distribution does not strongly affect the probability of outbreak occurrence or the overall size of large epidemics with crisis. This outcome is consistent with the expectation that fine-scale variability processes tend to average over long time scales and large host populations, exerting a limited influence on the overall size of large, sustained epidemics (36). Taken together, these results indicate that the choice of the EIP distribution primarily shapes the temporal structure of epidemics rather than their cumulative size, with potential implications for early epidemic growth and for the interpretation of short-term epidemic projections, including the timing and intensity of healthcare demand.

For total epidemic size, patterns differed in large epidemics without crisis, with differences between EIP assumptions that were less straightforward to interpret. This category encompassed a wide range of epidemic profiles and durations, likely increasing variability in model outcomes and reducing comparability between simulations. Such structural heterogeneity may therefore have amplified the apparent differences between EXP and ED assumptions. However, these differences could also reflect a genuine effect of the EIP distribution that depends on epidemic intensity. In this epidemic category, some outbreaks reached the cumulative infection threshold (>100 total human infections) over long simulation periods without showing a distinct epidemic peak, a pattern more frequent under the ED assumption. This observation can be explained by the interaction between the shape of the EIP distribution and stochastic transmission dynamics. Under the EXP assumption, a small fraction of mosquitoes becomes infectious very soon after infection, occasionally accelerating early transmission and allowing the epidemic to self-sustain. In contrast, the ED assumption produces a more synchronized and delayed onset of mosquito infectiousness, reducing the likelihood of rapid stochastic amplification and leading to smaller overall epidemic sizes in intermediate transmission regimes.

Our modelling framework also allowed us to explore how the differences between assumptions varied with parameter values, particularly for large epidemics with crisis. The largest variations were associated with the adult mosquito mortality rate, higher mortality generally corresponding to larger differences between EXP and ED assumptions. Consistent with the mechanisms described above, the EXP distribution allowed a fraction of mosquitoes to become infectious earlier, whereas the ED distribution delayed the onset of infectiousness. When mosquito mortality was low, these differences tended to remain limited, but as mortality increased, a substantial fraction of mosquitoes died before completing the mean EIP, amplifying the divergence between assumptions. This finding has practical implications for vector-borne disease management, particularly when control measures rely on vector population reduction (10), since predictions from models assuming an exponential EIP could potentially overestimate the transmission potential under such conditions. While such overestimation may be acceptable in modelling approaches that deliberately adopt conservative or worst-case assumptions to avoid underestimating risk, it could bias projections of intervention effectiveness or epidemic timing when accurate short-term dynamics are required. The human recovery rate was the second most influential parameter. Increasing the recovery rate reduced the differences between the two assumptions, suggesting that when the recovery rate increases, transmission dynamics may become more strongly governed by human rather than vector parameters, thus dampening the effect of EIP assumptions. The effects of the parameters related to the force of infection (biting rate and mosquito-to-human transmission probability) and the mean mosquito-to-human ratio ($\delta$) were less pronounced but remained detectable, especially for peak slope. Although the effects were weaker, differences between assumptions tended to increase with higher transmission intensity, particularly for peak size and peak slope. The wide range of values explored for all parameters was chosen to encompass diverse epidemiological contexts and to reflect the variability reported in previous studies (6,7). This variability is partly due to the difficulty of estimating these parameters (21) and to their spatial and temporal fluctuations (21,25). Therefore, assuming an exponential EIP distribution could have a varying impact depending on factors that influence the mosquito-to-human ratio and the biting rate, such as temperature, seasonality and rainfall (28).

One limitation of our study is that we used a single dataset to characterize the experimentally derived intra-vector distribution, representing one specific dengue-*Aedes aegypti* system. Although *Ae. aegypti* is one of the primary dengue vectors (37), conclusions may differ with other vector species or viral strains. One objective of this work was to provide a reusable and adaptable framework that can be applied to other scenarios by adjusting entomological, viral, and transmission parameters, as well as the epidemic thresholds used to define large epidemics and crises. While this modelling framework could be used to test and identify efficient control measures for managing arboviral transmission, such applications were not pursued here, as our study aimed to broadly examine epidemic dynamics under different EIP distributions rather than focus on specific intervention contexts.

The thresholds used to define large epidemics and crises may appear somewhat arbitrary, even though they were informed by previous studies for large epidemic threshold (20) and consistent with literature on dengue for the crisis threshold (30). These thresholds were chosen to create broad and interpretable categories, thereby minimizing the overlap between epidemic outcomes and facilitating comparative analyses across the large range of simulated scenarios. Importantly, they are not intended to be universal: they can be adjusted depending on the specific context to which the model is applied, including local epidemiological conditions and the capacity of health systems to manage symptomatic cases (e.g., hospital bed availability, staffing levels, and economic resources). The proposed framework is flexible and can be adapted to specific contexts, even if this was beyond the scope of the present study.

Some assumptions underlying our models also deserve discussion. First, while this study focused on the EIP, other infection-related durations, such as the human infectious period or intrinsic incubation period, have also been shown to influence epidemic dynamics (13,17). Here, we deliberately only varied the intra-vector distribution to isolate its specific impact, but future work could jointly explore the combined effects of realistic human and mosquito infection-period distributions.

Another assumption of our model concerns mosquitoes that fail to transition from the susceptible state ($S_M$) to the infected-disseminated ($ID_M$) state after taking an infectious blood meal. In the model, these mosquitoes returned susceptible and could become infected again during subsequent bites. This assumption is common in epidemiological transmission models (6), but its biological validity can be questioned. Because vector competence experiments sacrifice mosquitoes at fixed time points to assess infection, dissemination, and transmission statuses, it remains unclear whether mosquitoes that did not become infectious after being exposed could nevertheless experience barrier crossing following a subsequent exposure. Given the dose dependence between infection and dissemination barriers (38), it is plausible that a mosquito in which the virus initially fails to cross these barriers could nevertheless become infected after a later blood meal with a higher viral dose. Exploring alternative formulations, such as preventing reinfection attempts once a mosquito has failed infection once, could therefore provide additional insights into how this process influences transmission variability.

More generally, our findings should be interpreted as conditional on the modelling framework and assumptions adopted in this study. In particular, this includes the formulation of the force of infection, adapted to account for temporal variations in mosquito abundance. Alternative vector–host transmission models, incorporating different representations of transmission processes or heterogeneities, could lead to quantitatively different outcomes. Assessing the robustness of the effects reported here across alternative model structures and assumptions therefore remains a direction for future work.

Overall, this study proposes a modelling framework that integrates experimentally jointly derived information on the host-to-vector transmission parameter and on the distribution of the extrinsic incubation period into a dengue transmission model. This represents a step toward improving the biological realism and predictive accuracy of vector-borne disease models, thereby supporting the design of more efficient control strategies against arboviruses.

## AKNOWLEDGEMENT

We thank Anne Lehebel (BIOEPAR, Nantes) for her valuable advice and insightful revisions of the statistical analyses. All computations were run on the MIGALE computing cluster

**SUPPLEMENTARY MATERIAL**

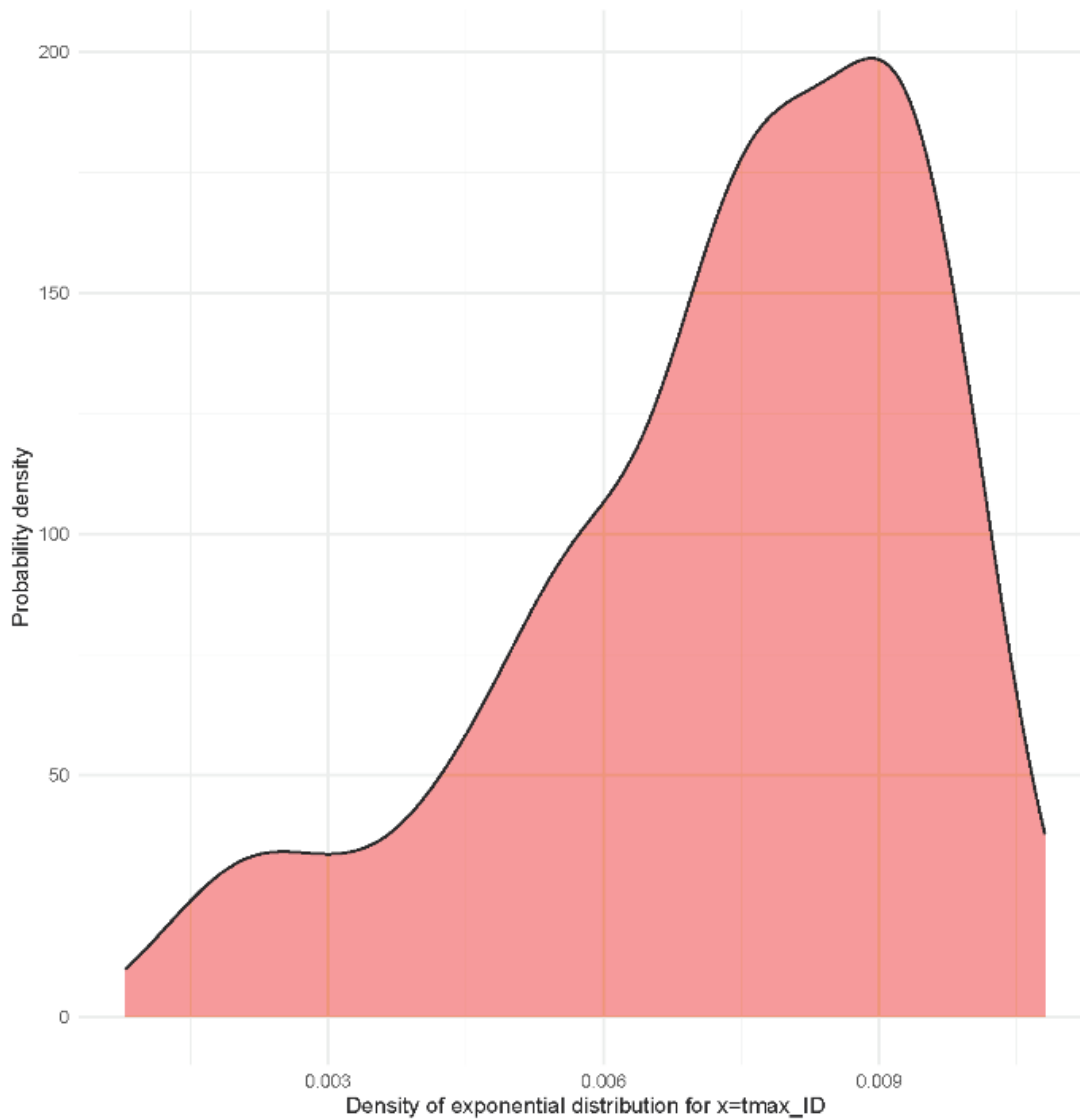

**Figure S1: Assessment of the exponential approximation and truncation used to approximate experimentally derived EIP distribution.** This figure evaluates the impact of truncating the exponential distribution used to approximate the mean duration spent in the infected–disseminated mosquito state ($ID_M$). For each IVD parameter set, the exponential distribution is parameterised using the mean duration derived from the Beta distributions describing the infected (I) and disseminated (D) states. The plot shows the distribution across parameter sets of the exponential density evaluated at the truncation time t = tmax_ID, providing an indication of the fraction of density remaining near the truncation boundary.

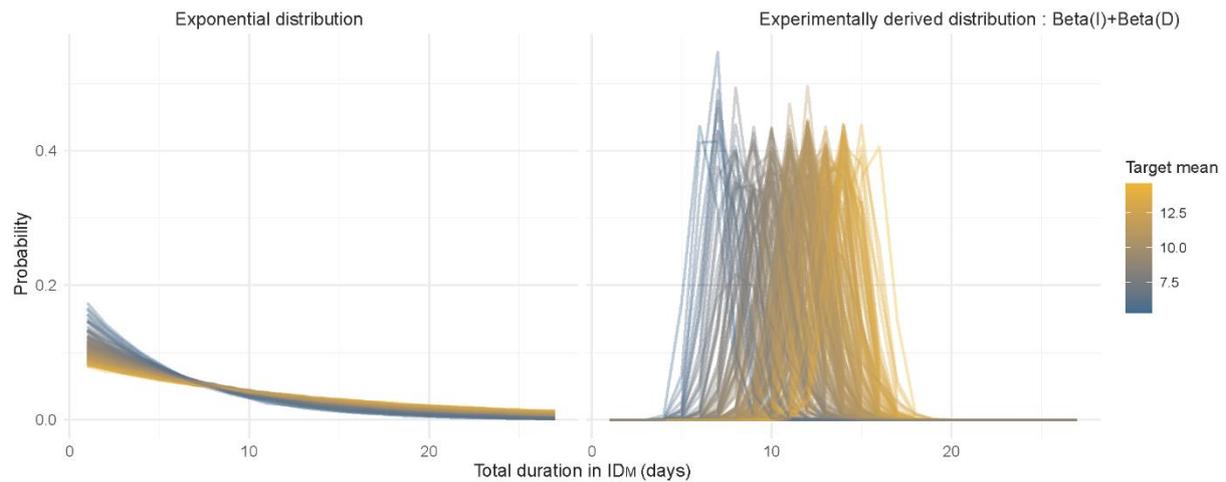

**Figure S2: Reconstructed distributions of the duration spent in the infected–disseminated mosquito state under experimentally derived and exponential distribution.** Each curve represents the distribution of the total duration (in days) spent in the infected–disseminated mosquito state. Durations are discretised at a daily time step, and probabilities indicate the proportion of mosquitoes experiencing a given residence time. Distributions were reconstructed from a random sample of 200 particles drawn from the set of particles used in the dengue transmission model. The exponential distributions are parameterised to match the mean duration of the experimentally derived distributions, so that differences between panels reflect differences in distribution shape rather than in mean residence time. Colours indicate the target mean duration associated with each particle; the same colour is used for the experimentally derived and exponential distributions corresponding to a given particle.

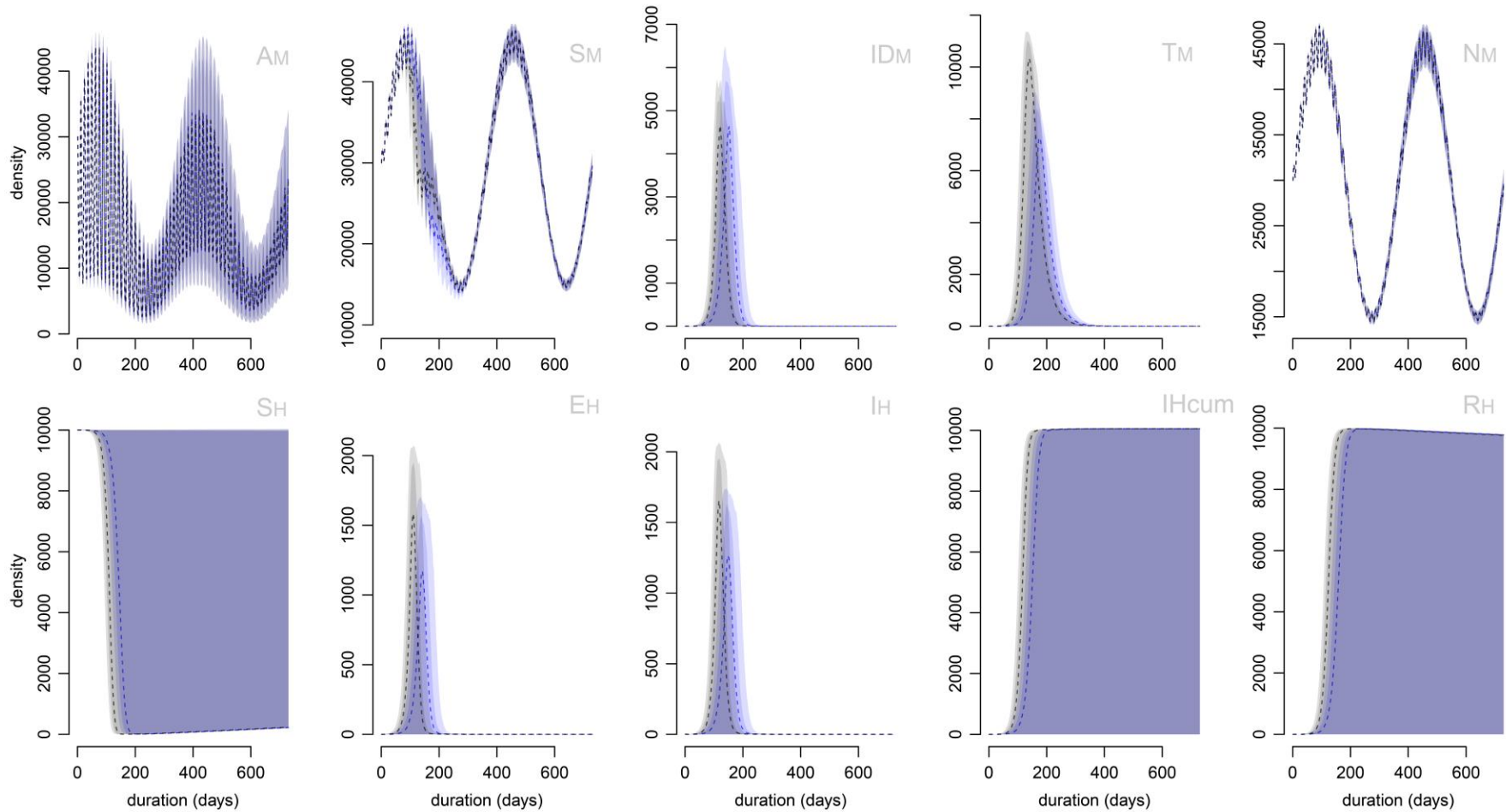

**Figure S3: Example of simulated model outputs for a single parameter set under the exponential (EXP) and experimentally derived (ED) EIP formulations.**
Each panel shows the temporal evolution (over 730 days) of one model state variable across 500 stochastic runs for the same parameter set. Mosquitoes are represented by an aquatic stage ($A_M$), and adult female compartments including susceptible ($S_M$), infected–disseminated ($ID_M$) and infectious ($T_M$) mosquitoes ($N_M = S_M + ID_M + T_M$); while humans are divided into susceptible ($S_H$)), exposed ($E_H$), infectious ($I_H$), and recovered ($R_H$) compartments ($N_H = S_H + E_H + I_H + R_H$). For each formulation, the dashed line represents the median (q50) trajectory, while shaded envelopes show simulation variability: the dark envelope corresponds to the interquartile range (q25–q75) and the lighter envelope to the 90% interval (q05–q95). The simulations were generated using the following parameter values: $N_H(0) = 10000$, $I_H(0) = 1$, $\omega = 8$ days$^{-1}$, $\mu_A = 0.1$ days$^{-1}$, $\tau = 0.15$ days$^{-1}$, $\mu_M = 0.033$ days$^{-1}$, $\nu = 0.4$ days$^{-1}$, $\beta_{mh} = 0.5$, $\mu_H = 0.000046$ days$^{-1}$, $\rho = 0.125$ days$^{-1}$, $\sigma = 0.14$ days$^{-1}$, $\delta = 3$.

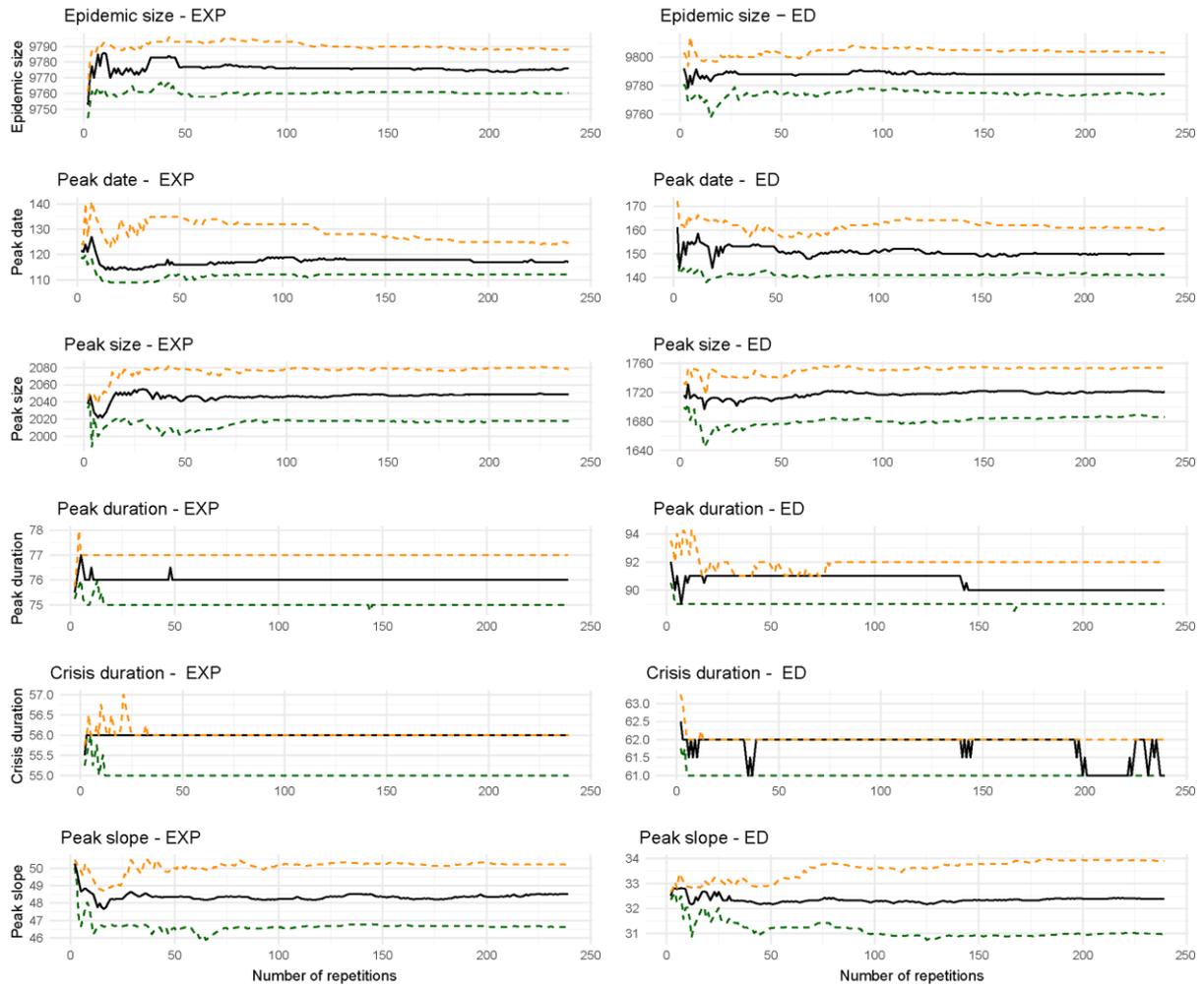

**Figure S4: Evolution of percentile estimates with the number of simulations (large epidemics with crisis).**

For each indicator (cumulative epidemic size, peak time, peak size, 10% peak duration, crisis duration, and peak slope) figures show the variations of the 25th (green), 50th (black), and 75th (orange) percentiles with the increase in the number of simulation repetitions. Results are displayed for the exponential intra-vector distribution assumption (EXP, left column) and the experimentally derived intra-vector distribution assumption (ED, right column). "Large epidemic" runs are defined by total human infections ≥ 100. "With crisis" runs are those where the number of concurrently infectious individuals reached at least 500 at any time ($I_H \geq 500$). Percentile curves stabilize around 100 repetitions. Based on this, we selected parameter combinations which led to large epidemic with crisis only if ≥100 of its simulations produced large epidemics with crisis; to meet this requirement we performed 500 simulations per particle. The simulations were generated using the following parameter values: $N_H(0) = 10000$, $I_H(0) = 1$, $\omega = 8$ days$^{-1}$, $\mu_A = 0.1$ days$^{-1}$, $\tau = 0.15$ days$^{-1}$, $\mu_M = 0.033$ days$^{-1}$, $\nu = 0.4$ days$^{-1}$, $\beta_{mh} = 0.5$, $\mu_H = 0.000046$ days$^{-1}$, $\rho = 0.125$ days$^{-1}$, $\sigma = 0.14$ days$^{-1}$, $\delta = 3$.

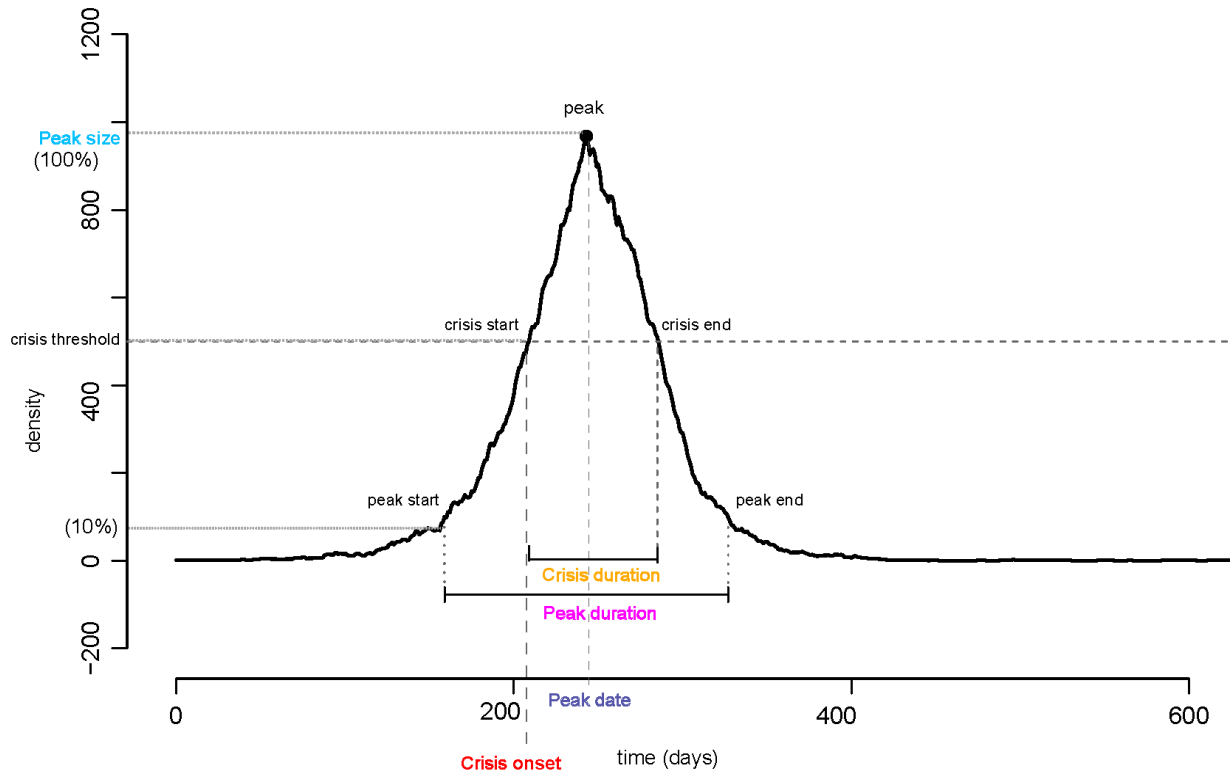

**Figure S5: Schematic representation of epidemic indicators used to characterise human epidemic dynamics.**

The figure illustrates how epidemic indicators were derived from simulated human infection trajectories, including peak size and date, as well as peak and crisis durations. Crisis onset and end were defined using a fixed threshold on the number of simultaneously infectious individuals, allowing the classification of epidemics according to their magnitude, timing, and duration.

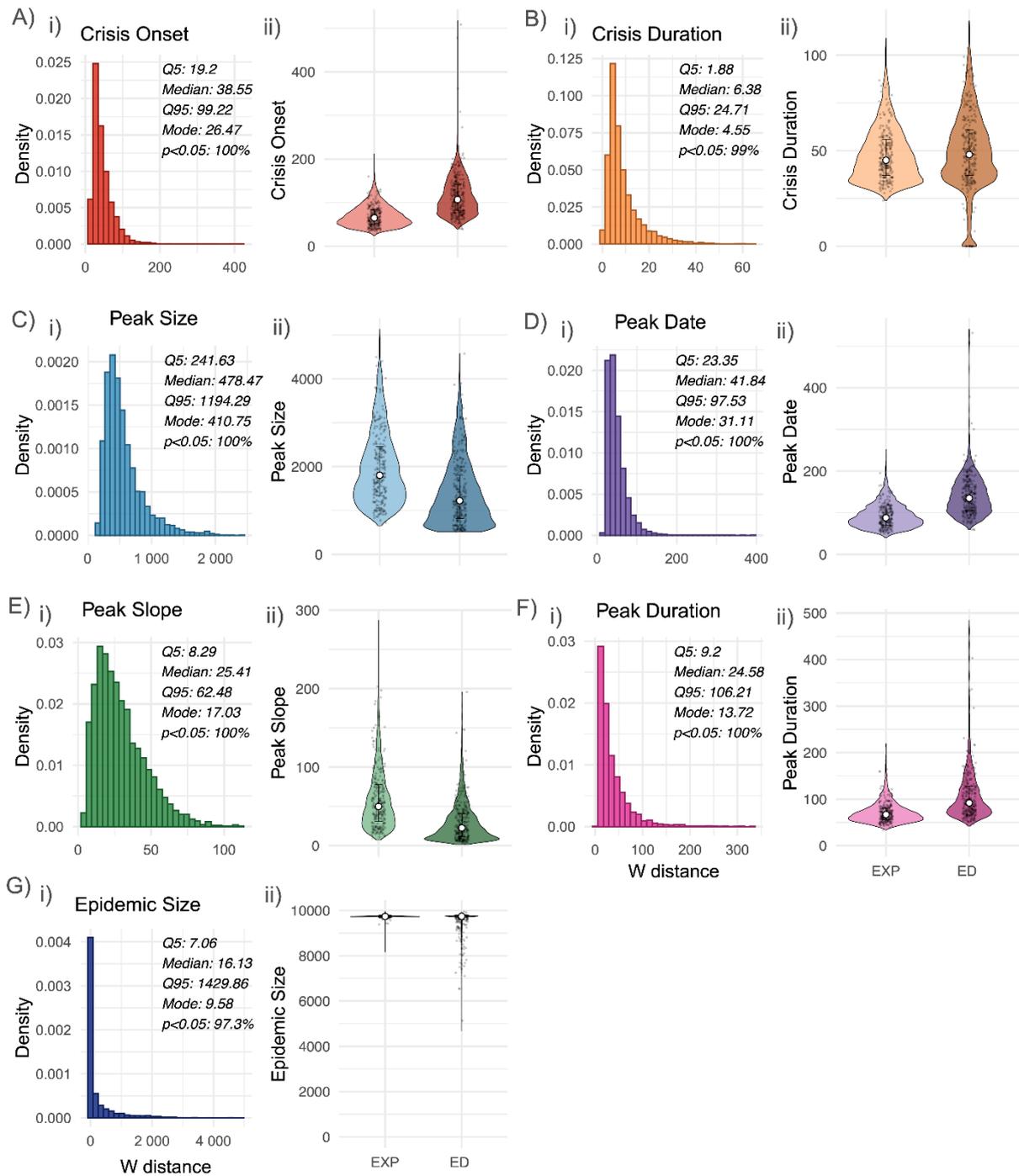

**Figure S6: Differences in epidemic indicators (large epidemics with crisis) when using an exponential (EXP) *vs.* an experimentally derived (ED) extrinsic incubation period (EIP) distribution evaluated with Wasserstein (W) distance**. (A) Crisis onset (first day when the number of concurrently infectious humans reached the "crisis" threshold, $I_H \geq 500$), (B) Crisis duration (time interval between the first and last day when $I_H \geq 500$). (C) peak size ($I_{H_{max}}$, maximum number of concurrently infectious humans), (D) peak date (time of $I_{H_{max}}$), (E) peak slope (average daily increase in $I_H$ between 10 % of the peak size and the peak size), (F) peak duration (time interval between the first and last day when $I_H$ exceeded 10 % of the peak size) and (G) total epidemic size (final cumulative number of human infections over the two-year simulation period). Each epidemic indicator is summarized by: (i) a histogram of the Wasserstein distance between epidemic indicator for ED and EXP computed per parameters combination*; (ii) violin plots showing the distribution of the medians
27

of epidemic indicators for each parameters combination for each assumption (small black dots: sample of medians (500 max); horizontal bar: Q25–Q75 interquartile range; white–black circle: median). Insets in (i) report quantiles, mode, and the proportion of parameters combination with statistically significant differences between assumptions (Wilcoxon rank-sum test, $p < 0.05$). *Parameter combinations were generated using a Latin Hypercube Sampling (LHS) design, in which parameter values were simultaneously varied across their literature-based ranges to produce 12,000 combinations. Analysis was restricted to parameter combinations that produced at least 100 large epidemics with crisis under each assumption (i.e. total number of human infections ≥ 100 and at least 500 concurrently infectious individuals), resulting in 3952 parameter combinations used.*

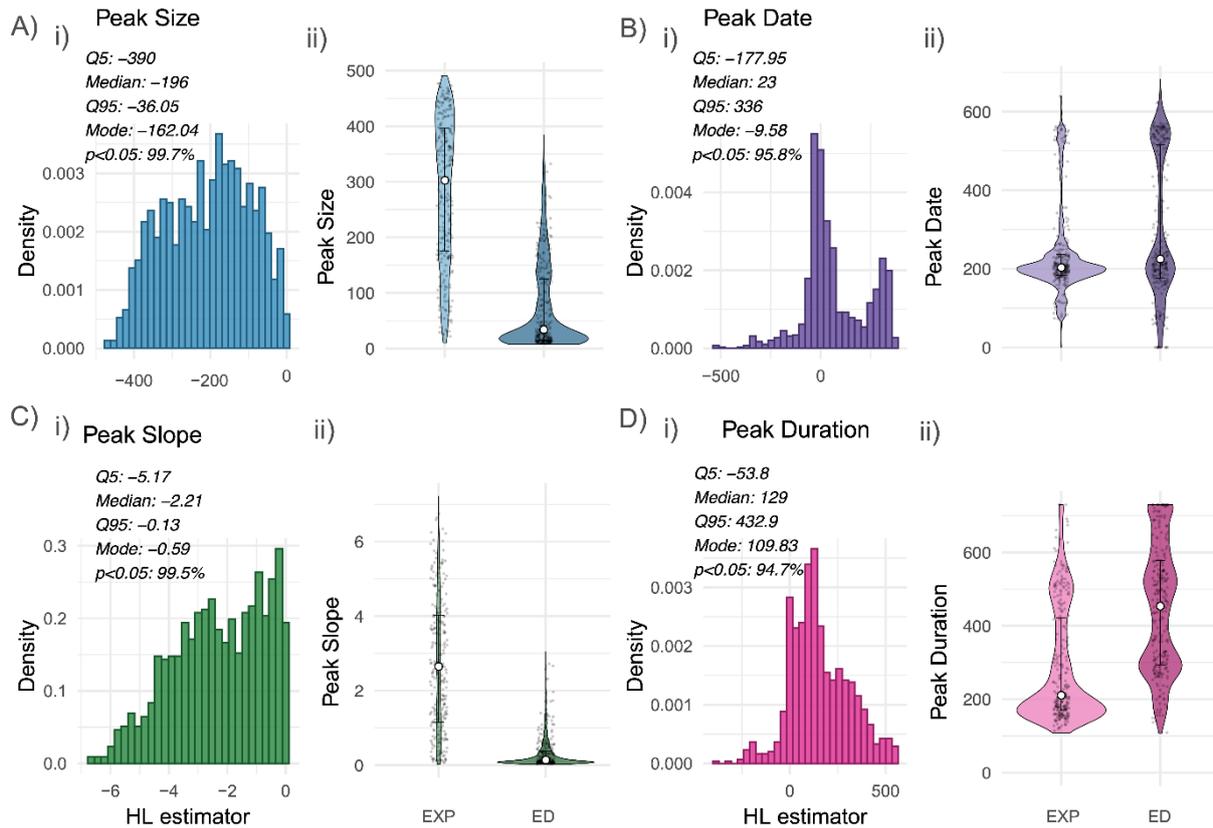

**Figure S7: Differences in epidemic indicators (large epidemics without crisis) when using an exponential (EXP) *vs.* an experimentally derived (ED) extrinsic incubation period (EIP) distribution evaluated with Hodges-Lehmann (HL) estimator.** (A) Peak size ($I_{H_{max}}$, maximum number of concurrently infectious humans), (B) peak date (time of $I_{H_{max}}$), (C) peak slope (average daily increase in $I_H$ between 10 % of the peak size and the peak size), and (D) peak duration (time interval between the first and last day when $I_H$ exceeded 10 % of the peak size). Each epidemic indicator is summarized by: (i) a histogram of the Hodges-Lehmann (HL) estimator of the difference between epidemic indicator for ED and EXP EIP computed per parameter combination* (negative value: larger indicators with EXP EIP); (ii) violin plots showing the distribution of the medians of epidemic indicators for each parameter combination for each assumption (small black dots: sample of medians (500 max); horizontal bar: Q25-Q75 interquartile range; white–black circle: median). Insets in (i) report quantiles, mode, and the proportion of parameter combinations with statistically significant differences between EXP and ED assumptions (Wilcoxon rank-sum test, p-value < 0.05). * *Parameter combinations were generated using a Latin Hypercube Sampling (LHS) design, in which model parameters were simultaneously varied across their literature-based ranges to produce a total of 12,000 parameter combinations. Analysis was restricted to parameter combinations that produced large epidemics without crisis under each assumption (i.e. total number of human infections ≥ 100 and fewer than 500 concurrently infectious individuals at any time), resulting in 942 parameter combinations used.*

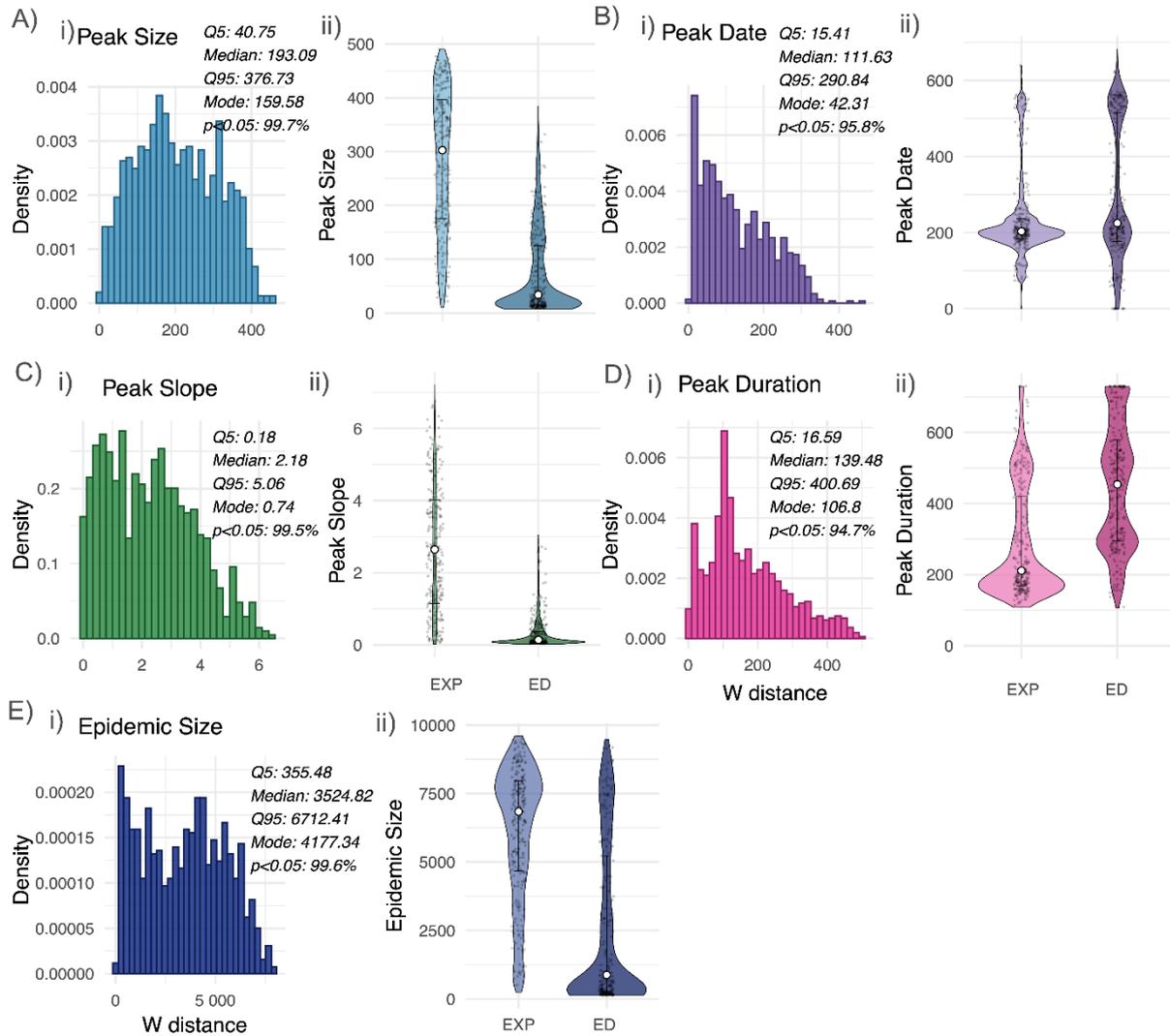

**Figure S8. Differences in epidemic indicators (large epidemics without crisis) when using an exponential (EXP) *vs.* an experimentally derived (ED) extrinsic incubation period (EIP) distribution evaluated using Wasserstein (W) distance**.

(A) peak size ($I_{H max}$, maximum number of concurrently infectious humans), (B) peak date (time of $I_{H max}$), (C) peak slope (average daily increase in $I_H$ between 10 % of the peak size and the peak size), (D) peak duration (time interval between the first and last day when $I_H$ exceeded 10 % of the peak size) and (E) total epidemic size (final cumulative number of human infections over the two-year simulation period). Each epidemic indicator is summarized by: (i) a histogram of the Wasserstein distance between epidemic indicators for ED and EXP computed per parameters combination* ; (ii) violin plots showing the distribution of the medians of epidemic indicators for each parameter combinations for each assumption (small black dots: sample of medians (500 max); horizontal bar: Q25–Q75 interquartile range; white–black circle: median). Insets in (i) report quantiles, mode, and the proportion of parameter combinations with statistically significant differences between assumptions (Wilcoxon rank-sum test, p < 0.05). *Parameter combinations were generated using a Latin Hypercube Sampling (LHS) design, in which model parameters were simultaneously varied across their literature-based ranges to produce a total of 12,000 parameter combinations. Analysis was restricted to parameter combinations that produced large epidemics without crisis under each assumption (i.e. total number of human infections ≥ 100 and fewer than 500 concurrently infectious individuals at any time), resulting in 942 parameter combinations used.*

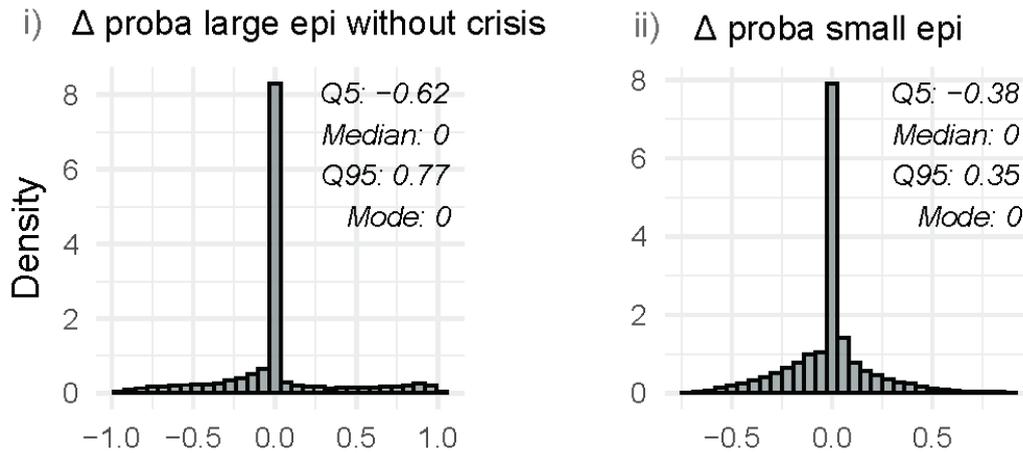

**Figure S9. Differences in probability of small epidemic and large epidemic without crisis occurrence when using an exponential (EXP) *vs.* an experimentally derived (ED) extrinsic incubation period (EIP) distribution.**

Histogram of the differences (ED – EXP) between probability that a parameter combination* lead to i) a large epidemic without crisis ($I_{Hcum} \geq 100$ and $I_H < 500$) and ii) small epidemics ($I_{Hcum} < 100$). Insets report quantiles, mode, and the proportion of parameter combinations with statistically significant differences between assumptions (Wilcoxon rank-sum test, $p < 0.05$). *Parameter combinations were generated using a Latin Hypercube Sampling (LHS) design, in which model parameters were simultaneously varied across their literature-based ranges to produce a total of 12,000 parameter combinations*

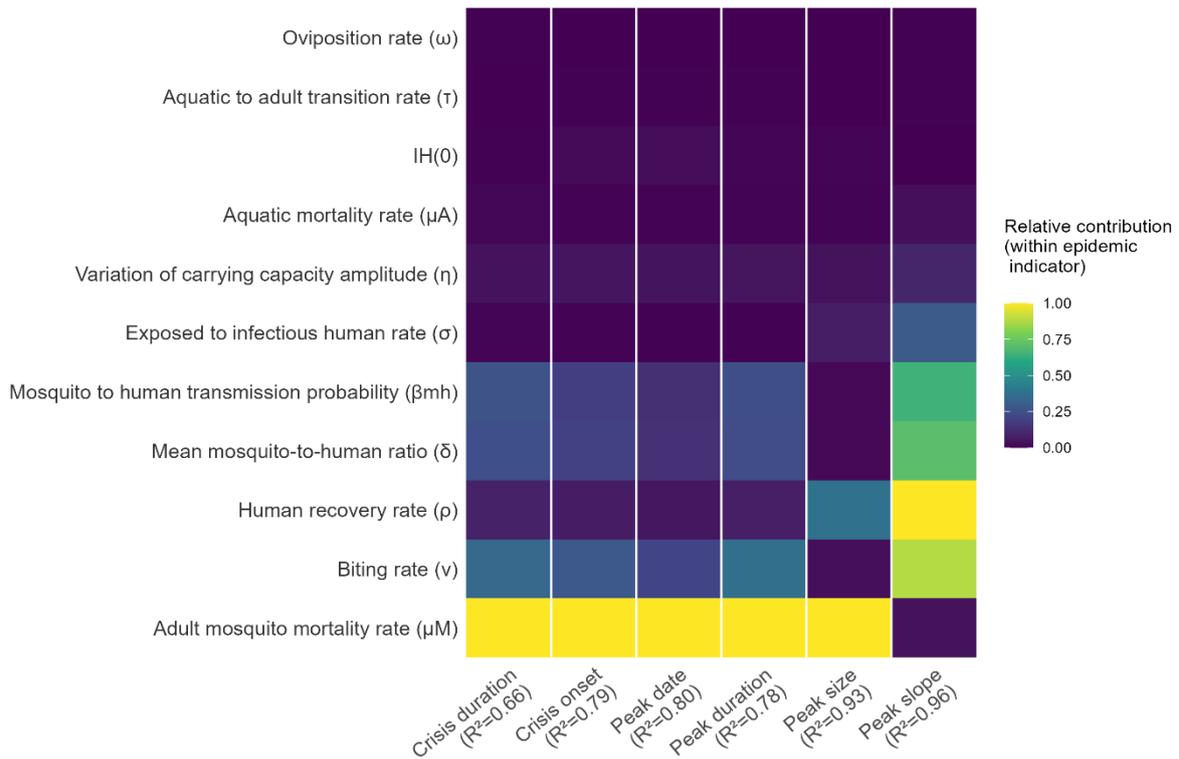

**Figure S10. Relative contribution of model parameters to differences between EXP and ED assumption for large epidemics with crisis estimated with the Wasserstein distance.**

Heatmap of ANOVA-based relative contributions of model parameters to the variability of the Wasserstein distance, computed across the Latin Hypercube Sampling (LHS) design restricted to parameter sets that generated at least 100 large epidemics with crisis. For each epidemic indicator, parameter contributions were derived from second-order linear models (including main effects and pairwise interactions) and aggregated per parameter. Colour intensities represent the relative contribution of each parameter to the variability of Wasserstein distance for a given epidemic indicator, rescaled to the [0,1] range within each indicator.

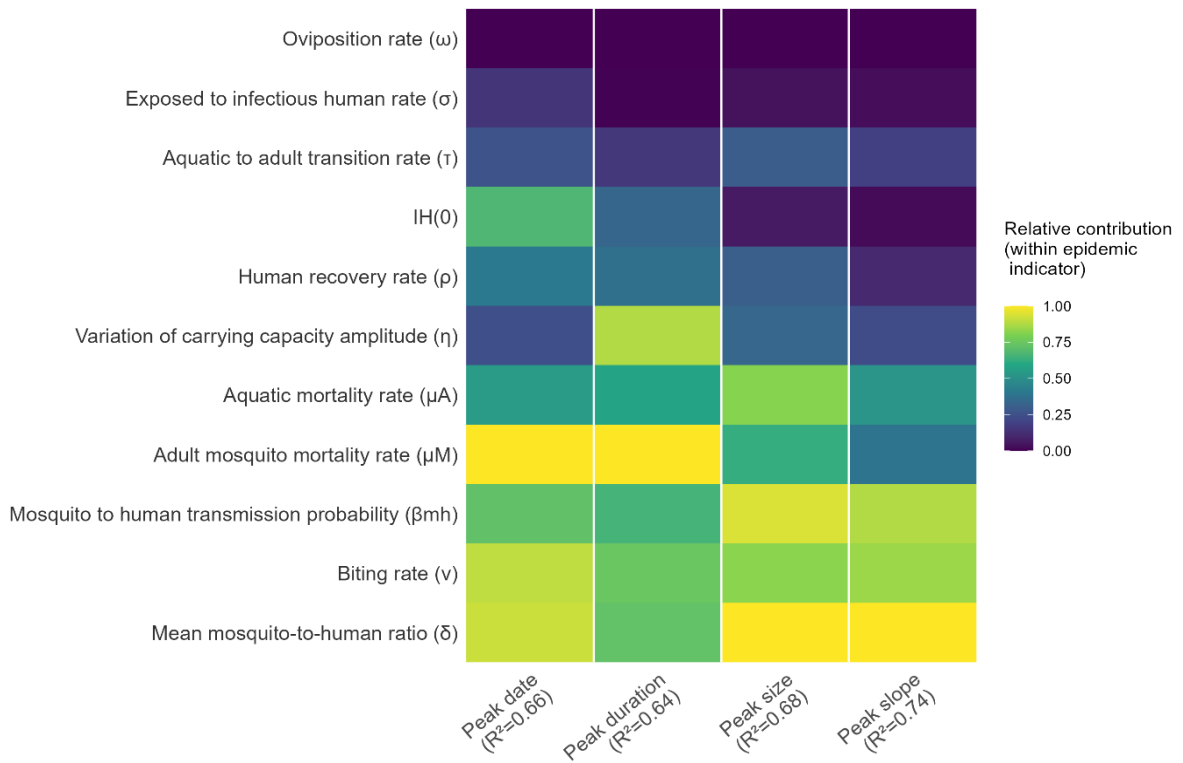

**Figure S11. Relative contribution of model parameters to differences between EXP and ED assumption for large epidemics without crisis estimated with the Hodges–Lehmann (HL) estimators.**

Heatmap of ANOVA-based relative contributions of model parameters to the variability of the Hodges–Lehmann (HL) estimators, computed across the Latin Hypercube Sampling (LHS) design restricted to parameter combinations that generated at least 100 large epidemics without crisis. For each epidemic indicator, parameter contributions were derived from second-order linear models (including main effects and pairwise interactions) and aggregated per parameter. Colour intensities represent the relative contribution of each parameter to the variability of HL for a given epidemic indicator, rescaled to the [0,1] range within each indicator.

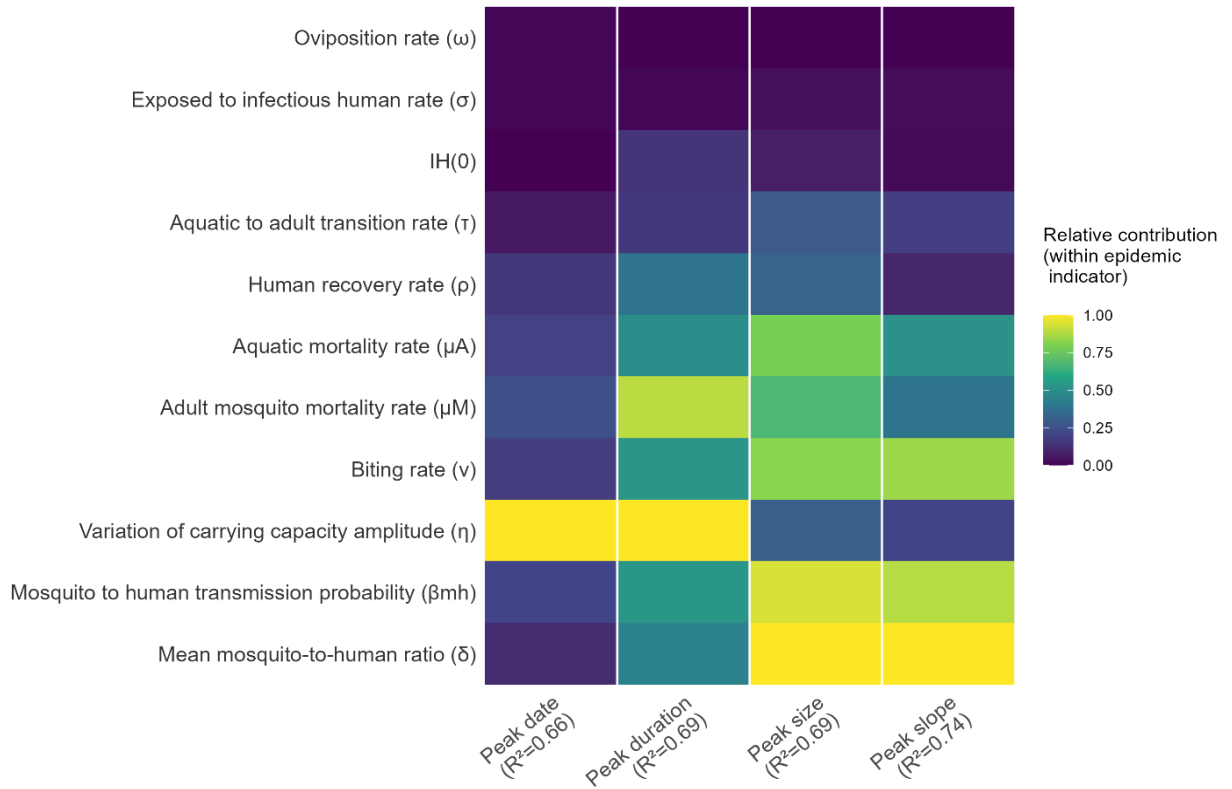

**Figure S12. Relative contribution of model parameters to differences between EXP and ED assumption for large epidemics without crisis estimated with the Wasserstein distance.**

Heatmap of ANOVA-based relative contributions of model parameters to the variability of the Wasserstein distance, computed across the Latin Hypercube Sampling (LHS) design restricted to parameter sets that generated at least 100 large epidemics without crisis. For each epidemic indicator, parameter contributions were derived from second-order linear models (including main effects and pairwise interactions) and aggregated per parameter. Colour intensities represent the relative contribution of each parameter to the variability of Wasserstein distance for a given epidemic indicator, rescaled to the [0,1] range within each indicator.

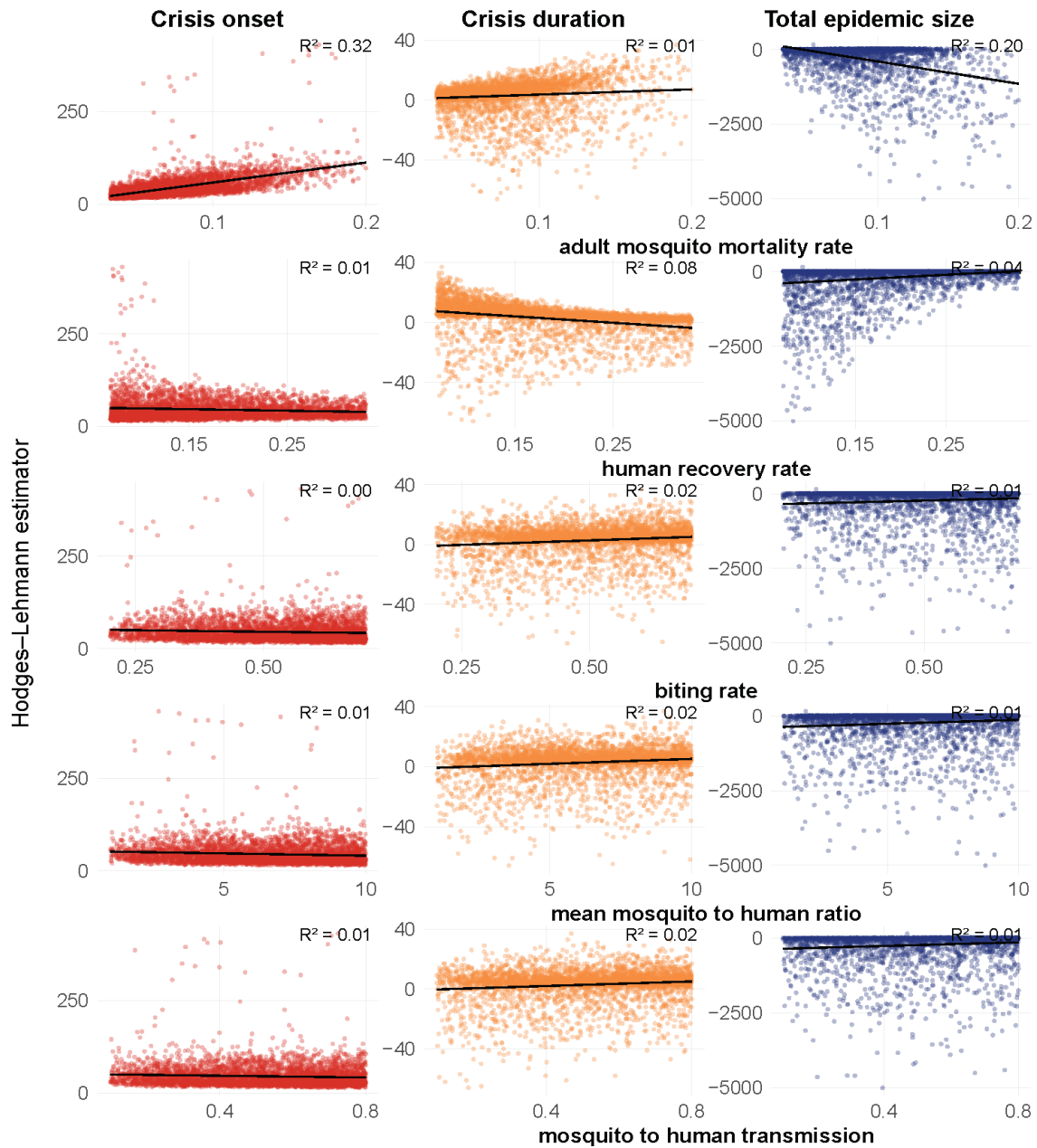

**Figure S13. Impact of parameters variability on differences between the EXP and ED assumptions for large epidemics with crisis.**

Each panel shows the marginal variation of the Hodges–Lehmann (HL) estimator with respect to one key parameter, while all other parameters vary across the Latin Hypercube Sampling design restricted to parameter combinations that generated large epidemics with crisis under each assumption. Columns correspond to epidemic indicators: (1) Crisis onset (first day when the number of concurrently infectious humans reached the "crisis" threshold, $I_H \geq 500$), (2) Crisis duration (time interval between the first and last day when $I_H \geq 500$), (3) total epidemic size (final cumulative number of human infections over the two-year simulation period). Rows correspond to parameters: (1) adult mosquito mortality rate ($\mu_M$), (2) human recovery rate ($\rho$), (3) biting rate ($\nu$), (4) mean mosquito-to-human ratio ($\delta$), (5) mosquito-to-human transmission probability ($\beta mh$). Points represent individual HL estimators, and black lines indicate average linear trends shown for visual guidance only and not intended for quantitative inference.

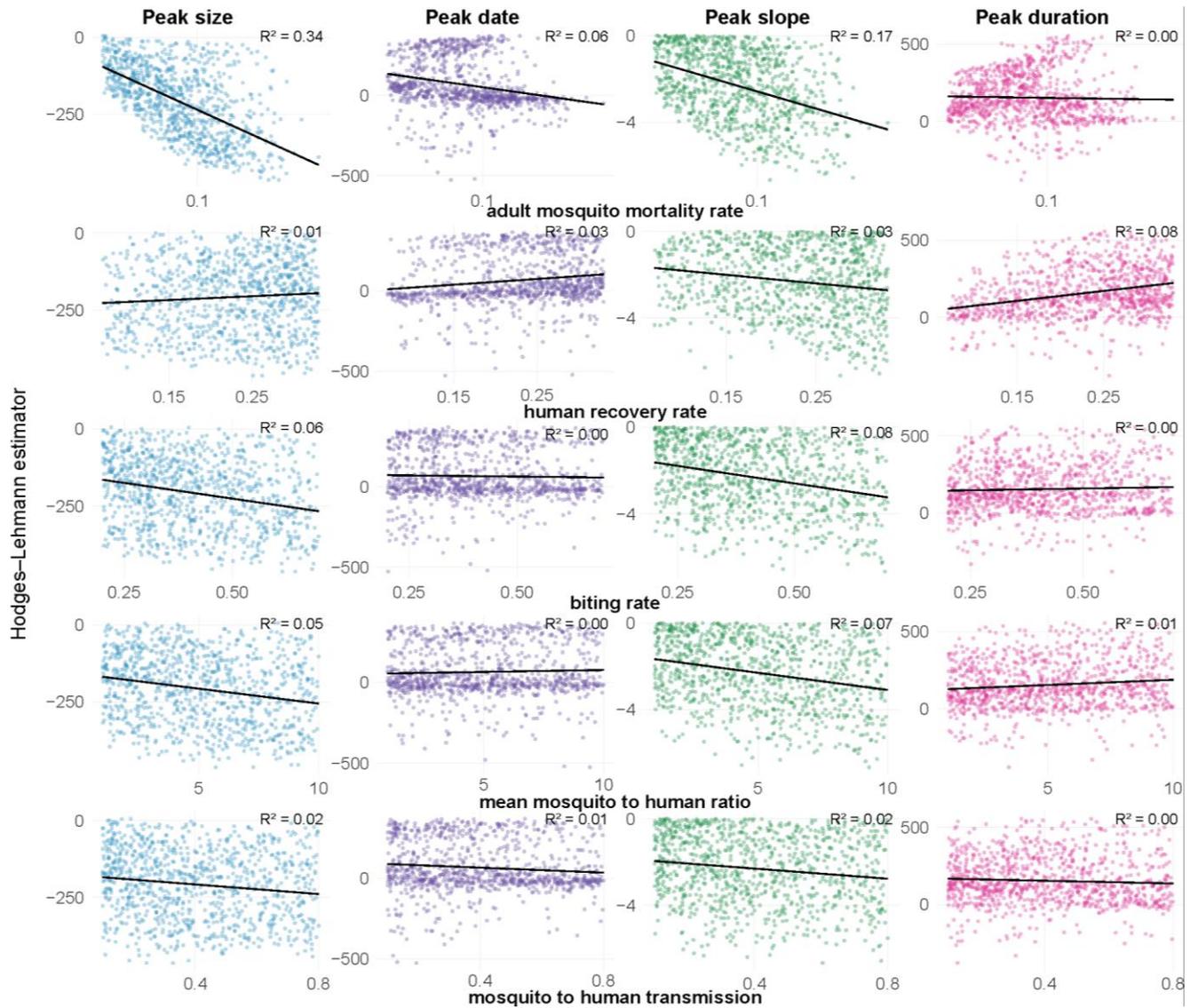

**Figure S14. Impact of key parameters on differences between the EXP and ED assumptions for large epidemics without crisis.**

Each panel shows the marginal variation of the Hodges–Lehmann (HL) estimator with respect to one key parameter, while all other parameters vary across the Latin Hypercube Sampling design restricted to parameter combinations that generated at least 100 large epidemics without crisis under each assumption. Columns correspond to epidemic indicators: (1) peak size ($I_{H_{max}}$, maximum number of concurrently infectious humans), (2) peak date (time of $I_{H_{max}}$), and (3) peak slope (average daily increase in $I_H$ between 10 % of the peak size and the peak size), and (4) peak duration (time interval between the first and last day when $I_H$ exceeded 10% of the peak value). Rows correspond to parameters: (1) adult mosquito mortality rate ($\mu_M$), (2) human recovery rate ($\rho$), (3) biting rate ($\nu$), (4) mean mosquito-to-human ratio ($\delta$) and (5) mosquito-to-human transmission probability ($\beta_{mh}$). Points represent individual HL estimators, and black lines indicate average linear trends shown for visual guidance only and not intended for quantitative inference.